\documentclass[aps,pra,superscriptaddress,preprint]{revtex4-1}
\usepackage{graphicx}  
\usepackage{bm}        
\usepackage{amssymb}   
\usepackage{natbib}
\usepackage{amsmath}
\usepackage[colorlinks=true,citecolor=red,linkcolor=blue]{hyperref}
\begin{document}
\title{Plasmon transmission through excitonic subwavelength gaps}

\author{Maxim Sukharev}
\email{maxim.sukharev@asu.edu}
\affiliation{Science and Mathematics Faculty, College of Letters and Sciences, Arizona State University, Mesa, Arizona 85212, USA}

\author{Abraham Nitzan}
\email{anitzan@upenn.edu}
\affiliation{School of Chemistry, Tel Aviv University, Tel Aviv 69978, Israel}
\affiliation{Department of Chemistry, University of Pennsylvania, Philadelphia, Pennsylvania, 19104, USA}

\date{\today}
\begin{abstract}
We study the transfer of electromagnetic energy across a subwavelength gap separating two co-axial metal nanorodes. The absence of spacer in the gap separating the rods the system exhibits the strong coupling between longitudinal plasmons in the two rods. The nature and magnitude of this coupling is studied by varying various geometrical parameters. When the length of one rod is varied this mode spectrum exhibits the familiar anti-crossing behavior that depends on the coupling strength determined by the gap width. As a function of frequency the transmission is dominated by a splitted longitudinal plasmon peak. The two hybrid modes are the dipole-like "bonding" mode characterized by a peak intensity in the gap, and a quadrupole-like "antibonding" mode whose amplitude vanishes at the gap center. When off-resonant $2-$level emitters are placed in the gap, almost no effect on the frequency dependent transmission is observed. In contrast, when the molecular system is resonant with the plasmonic lineshape, the transmission is strongly modified, showing characteristics of strong exciton-plasmon coupling, modifying mostly the transmission near the lower frequency "bonding" plasmon mode. The presence of resonant molecules in the gap affects not only the molecule-field interaction but also the spatial distribution of the field intensity and the electromagnetic energy flux across the junction.
\end{abstract}
\maketitle

\section{Introduction}
 \label{introduction}
Studies aimed at understanding the consequences of the interaction of electromagnetic fields with metal, semiconductor and molecular nanostructures under the effort to construct, characterize, manipulate and control plasmonic devices \cite{maradudin2014modern}. Recurring themes in these studies are the plasmonic response of aggregates of nano-particles \cite{Nikoobakht:2003aa,Gantzounis:2005aa,Talley:2005aa,imura2006near,Jain:2006aa,Jain:2007aa,0953-8984-19-9-096210,Jain:2008aa,Li:2009aa,Chergui:2009aa,Miljkovic:2010aa,Wustholz:2010aa,Farcau:2010aa,Letnes:2011aa,Encina:2011aa,Whitmore:2011aa,Jung:2011aa,Banik:2012aa,Chang:2012aa,Banik:2012ab,Fraire:2013aa,Prodan419,Nordlander:2004aa,Nordlander:2004ab,Prodan:2004aa,Halas:2011aa,Esteban:2012aa,doi:10.1021/nl4014887,Harris:2009aa}, and the possibility to transmit electromagnetic energy over constrictions substantially smaller than the radiation wavelength \footnote{other important recurring themes such as induction of transient dielectric properties, lifetime of radiative and non-radiative processes, and formation and utilization of hot electrons and local hotspots are not related to the present study}. Composite structures (metal-dielectric, metal-semiconductor) are often found useful because light can localize at their interfaces. Of particular importance are metal- molecule composites where strong plasmon-exciton coupling together with the non-linear optical response of the molecular system can generate new physical behavior on one hand and provide more control capabilities by tuning the molecular subsystem on the other.

In this paper we study a class of systems of the latter kind, focusing on the light transmission properties of a model system comprising metal rods and molecular aggregates of sub-wavelength dimensions. Transmission properties of nano-size structure are usually discussed in two connotations. First, following early studies by Gersten and Nitzan \cite{Gersten:1984aa,Hua:1985aa} there is substantial interest in the way plasmonic particles affect excitation energy transfer in molecular systems \cite{Andrew1002,Gersten:2007aa,Zhang:2007aa,Reil:2008aa,Marocico:2009aa,Saini:2009aa,Chung2010,Martin-Cano:2010aa,Su:10,Faessler:2011aa,Marocico:2011aa,Pustovit:2011aa,Anton:2012aa,Lunz:2012aa,Zhao:2012aa,C2CP44010E,Karanikolas:2014aa,chou2015forster,4939206,Kucherenko2015,2040-8986-16-11-114015,Roslyak:2014aa}. Second, plasmonic structure can operate as sub-wavelength waveguides which are important in the construction of light controlled nano-devices and nano-size optical communication and information storage systems \cite{Barnes:2003aa,Maier:2003aa,Zia:04,Oulton:2008aa,Grandidier:2009aa,Sorger:2011aa,3604014,Ellenbogen:2011aa,Paul:2012aa,Solis:2013aa,gu2015photon,Carmeli:2015aa}. For the latter systems two principal design systems have been studied: one comprises a chain of nanoparticles \cite{Paul:2012aa,Sorger:2011aa}, where waveguiding is achieved by plasmon-hopping between nanoparticles and energy transport may be regarded as motion along a plasmonic band. The other uses the optical properties of metal- dielectric (including metal-vacuum) interfaces \cite{Barnes:2003aa,Oulton:2008aa,Sorger:2011aa,3604014,Ellenbogen:2011aa}. Molecular aggregates were used in both design types. In Refs. \cite{Maier:2003aa,Paul:2012aa,Solis:2013aa} they are used as reporters for the distribution of electromagnetic energy along the waveguide, while in Refs. \cite{Grandidier:2009aa,3604014,Ellenbogen:2011aa,Carmeli:2015aa} they constitute an active constituent of the dielectric subsystem. Another way in which a molecular component can affect the operation in devices based on plasmon hybridization and hopping is in affecting the hybridization characteristics of the system. Obviously, the interactions between plasmonic excitations on different particles will depend on the molecular environment between them \cite{3167407,3541820}. When the molecular optical response is far from resonance with the relevant plasmonic frequencies this dependence may be accounted for by incorporating a host medium with a suitable dielectric constant. In the linear response regime a frequency dependent dielectric function can represent a molecular system in resonance with the plasmonic spectrum, however such a procedure may not properly account for the distinction between lifetime and dephasing relaxation processes. Alternatively, the molecular subsystem is described here explicitly and quantum mechanically using a variant of the procedure described in Ref. \cite{PhysRevA.84.043802} (See also Refs. \cite{Bonifacio:1975aa,PhysRevA.47.1247,Judkins:95,Hofmann:1999aa,Slavcheva:2002aa,Slavcheva:2004aa,Fratalocchi:2008aa,Andreasen:09}). This makes it possible to address the regime of \textit{strong exciton-plasmon coupling}, where the molecular species does not only modify the plasmon-plasmon coupling but becomes an active component of the transmitting system. We note in passing that the latter procedure can also describe the molecular system in the non-linear response regime \cite{doi:10.1021/nn4054528,jcp_chirps14,Sukharev:2015aa,Blake:2015aa}, although we do not address this regime in the present study.

The effect of strong exciton-plasmon coupling on the optical response properties of metal-molecules composites have been under active study for some time \cite{0034-4885-78-1-013901}. Here we focus on its manifestation in the electromagnetic energy- transmission properties of a system comprising two metal cylinders aligned along a common axis with a gap of variable length between them. Light is injected into the system using a source point dipole located at one end of the two metal-rod system, and the transmitted intensity is recorded at the other end. The effect of a molecular aggregate filling the gap between the metal cylinders is studied in order to elucidate the role of several key parameters that characterize the molecular species (a) Molecular density, (b) Molecular transition frequency and (c) Molecular lifetime and dephasing relaxation rates. We note that some issues related to this study were addressed in previous works. On the experimental side, Benner et al \cite{1367-2630-15-11-113014,Benner:2014aa} have studied plasmon transmission (and electronic conduction) across small constrictions between metal wires, highlighting the need to account for heating and thermal expansion effects in realistic experiments. Neuhauser and coworkers \cite{3167407,3541820} have placed a single molecule between two metal spheres in order to simulate its effect on the plasmon-plasmon coupling. While their single molecule model uses a similar density matrix description that can study the role of molecular relaxation processes, these issues where not explicitly addressed in these studies. Our earlier work \cite{PhysRevA.84.043802} and a recent work by Sadeghi \cite{4767653} have addressed the role of molecular dephasing in the coherent response of strongly coupled exciton-plasmon systems. Gu and coworkers \cite{gu2015photon} have recently considered the transmission of electromagnetic signal across a gap in a configuration similar to the one studied here, however without molecules. The present study is aimed to elucidate the way such transmitted energy is affected by strong exciton-plasmon coupling.

\section{Model}
 \label{model}
Our system consists of two metal cylinders of lengths $L_1$ and $L_2$ and equal diameters $D$, lying along a common axis and separated by a gap of width $\Delta L$ (see the inset of Fig. (\ref{fig1})). This gap can be bridged by a molecular aggregate of the same diameter - an assembly of two-level point objects whose optical response is described by the optical Bloch equations \cite{PhysRevA.84.043802}. For comparison, we also consider the corresponding system with a vacuum gap, and the system with no gap ($\Delta L$).

The dynamics of the electric, $\vec{E}$, and magnetic, $\vec{H}$, fields is simulated using classical Maxwell's equations
   \begin{subequations}
   \label{Maxwell}
    \begin{align}
     & \mu_0 \frac{\partial \vec{H}}{\partial t}  =  -\nabla \times \vec{E}, \\
     & \varepsilon_0\frac{\partial \vec{E}}{\partial t}  =  \nabla \times \vec{H} - \vec{J}, 
    \end{align}
  \end{subequations}
 where $\varepsilon_0$ and $\mu_0$ are the permittivity and the permeability of free space, respectively. The current source in the Ampere law (\ref{Maxwell}b), $\vec{J}$, corresponds to either the current density in spatial regions occupied by metal (Eq. (\ref{Drude_J}) below) or the macroscopic polarization current, $\vec{J}=\frac{\partial\vec{P}}{\partial t}$, in space filled with a molecular aggregate. 

The dispersion of metal is taken into account via Drude model with the dielectric constant of metal, $\varepsilon\left(\omega\right)$, in the form
 \begin{equation}
\label{Drude_epsilon}
 \varepsilon\left(\omega\right)=\varepsilon_r-\frac{\omega_p^2}{\omega^2-i\Gamma\omega},
\end{equation}
where $\Gamma$ is the damping parameter, $\omega_p$ is the bulk plasma frequency, and $\varepsilon_r$ is the high-frequency limit of the dielectric constant. For the range of frequencies considered in this work the following set of parameters was chosen to represent gold: $\varepsilon_r=9.5$, $\omega_p=8.95$ eV, and $\Gamma=0.069$ eV \cite{doi:10.1117/1.3001731}.
The corresponding current density in the metal region is evaluated according to the following equation \cite{PhysRevB.68.045415}
 \begin{equation}
\label{Drude_J}
 \frac{\partial\vec{J}}{\partial t}+\Gamma\vec{J}=\varepsilon_0\omega_p^2\vec{E}.
\end{equation}

The optical response of the molecular subsystem is simulated using rate equations for a two-level system driven by a local electric field $\vec{E}$ \cite{siegman1986university}
   \begin{subequations}
   \label{rate_equations}
    \begin{align}
&\frac{dn_1}{dt}-\gamma_{21}n_2=-\frac{1}{\hbar\Omega_0}\vec{E}\cdot\frac{\partial\vec{P}}{\partial t}, \\
&\frac{dn_2}{dt}+\gamma_{21}n_2=\frac{1}{\hbar\Omega_0}\vec{E}\cdot\frac{\partial\vec{P}}{\partial t}, \\
&\frac{\partial^2\vec{P}}{\partial t^2}+(\gamma_{21}+2\gamma_{d})\frac{\partial\vec{P}}{\partial t}+\Omega_0^2\vec{P}=-\sigma(n_2-n1)\vec{E},
    \end{align}
 \end{subequations}
where $n_1$ and $n_2$ ($n_1+n_2=n_0$, where $n_0$ is the number density of molecules) correspond to the populations of the ground and the excited molecular states, respectively, $\vec{P}$ is the macroscopic polarization, $\gamma_{21}$ is the radiationless decay rate of the excited state, $\gamma_{d}$ is the pure dephasing rate, and $\hbar\Omega_0$ is the energy separation of the molecular levels. The coupling constant $\sigma$ is given by [\onlinecite{PhysRevA.91.043835}]
   \begin{equation}
   \label{constants}
\sigma=\frac{2\Omega_0\mu_{12}^2}{3\hbar},
  \end{equation}
where $\mu_{12}$ is the transition dipole moment. Eqs. (\ref{rate_equations}) and (\ref{constants}) assume that the molecular optical response is isotropic. Consequently, the orientation of the local induced polarization is along the polarization of the local electric field. In the standard situation where the individual molecule does not respond isotropically, the results reported below correspond to the assumption that the distribution of molecular orientations is isotropic. This aspect of the model can be generalized in order to investigate the interesting possibility that molecular effects in plasmon transport can be affected by the molecular orientation. In this work we do not consider such effects.

The system of coupled equations (\ref{Maxwell}), (\ref{Drude_J}), and (\ref{rate_equations}) is solved numerically on a multi-processor computer \cite{PhysRevA.84.043802}. The space is discretized in accordance with FDTD algorithm \cite{taflove2005computational} in three dimensions. The spatial resolution of $\delta x=\delta y=\delta z=1$ nm is chosen to achieve numerical convergence and avoid staircase effects (artificial "\textit{hot}" spots in the local field due to discretization of curved surfaces in Cartesian coordinates). The time step is $\delta t=\delta x/(2c)=1.7$ as, where $c$ is the speed of light in vacuum. Open boundaries are simulated using convolutional perfectly matched layers (CPML) \cite{CPML_paper}. We found that for a system considered here the best results were achieved with $19$ CPML layers. The total simulation domain for all calculations is $181\times181\times321$.

We employ short-pulse method (SPM) \cite{PhysRevA.84.043802} to calculate linear response of the plasmonic/excitonic system. The time envelope of the probe incident pulse, $f\left(t\right)$, is taken in the form of the Blackman-Harris window
   \begin{equation}
   \label{Blackman-Harris}
f\left(t\le\tau\right)=\sum_{n=0}^{3}a_n \text{cos}\left(\frac{2\pi n t}{\tau}\right)
  \end{equation}
The pulse duration is denoted as $\tau$, other parameters are: $a_0=0.3532$, $a_1=-0.488$, $a_2=0.145$, and $a_3=-0.0102$. The results discussed in the next section are obtained for $\tau=0.36$ fs pulse duration leading to an incident pulse with nearly flat spectrum for frequencies between $1$ eV and $2$ eV.

Unless otherwise stated, for the calculations reported below we employ two cylindrical identical metal wires with dimensions $L_1=L_2=100$ nm, and $D=20$ nm with a small empty gap in between. The system is excited locally by a pointwise soft source (classical point dipole) of the functional form given by Eq. (\ref{Blackman-Harris}) that is placed $5$ nm from the rightmost wire. Simulations with the driving dipole directed along the $z$-axis results in significantly higher transmission compared to any transverse polarization and we have therefore used this driving polarization. The transmitted intensity in the $z$ direction, $\left | E_z\right |^2$, is detected on the opposite side also at a distance of $5$ nm from the leftmost wire at a given point on the grid. It should be noted that the utilization of SPM implies that we consider only elastic scattering. For a case of interacting wires without molecules it is obviously the case as the optical response of metal is treated linearly using Drude model. In case of molecules one has to keep the incident peak amplitude low enough such that the population of the molecular excited state is always significantly smaller than $1$. In the simulations discussed below this condition was carefully monitored at all times.

\section{Results and discussion}
\label{results}
\begin{figure}[t!]
\begin{center}
\includegraphics[width=0.9\textwidth]{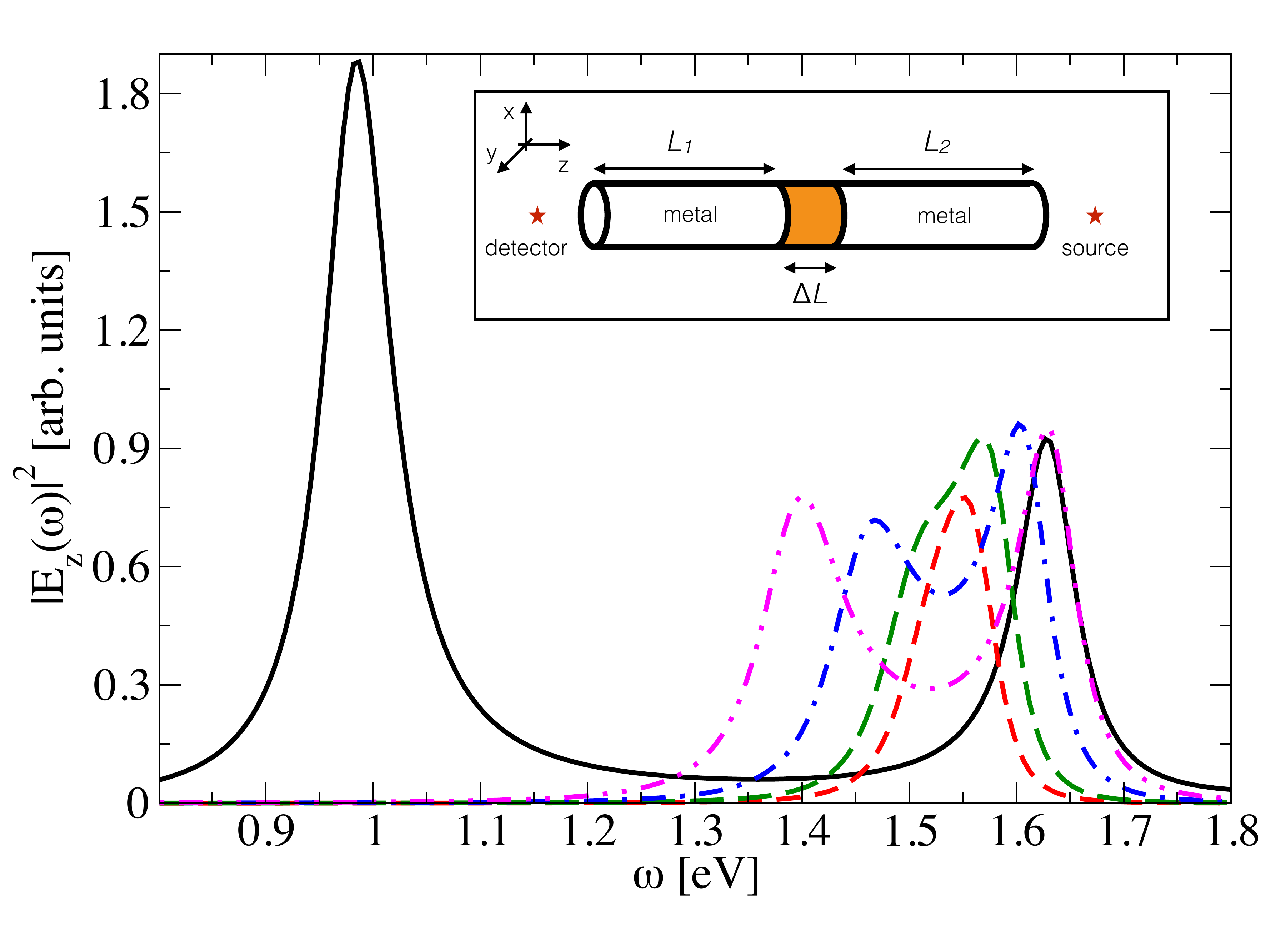}
\caption{\label{fig1} The inset schematically depicts two closely spaced wires with lengths $L_{1,2}$ separated by a subwavelength gap $\Delta L$. The excitation and detection are carried out locally at a distance of $5$ nm from the wires' surface with the source dipole oscillating in the $z$ direction. All simulations in this paper are carried out for wires with a circular cross-section of $20$ nm in diameter. The main panel shows the transmitted intensity in the longitudinal direction, $\left | E_z(\omega)\right |^2$, as a function of the incident frequency. The black line shows data for a single wire with $L_1=210$ nm. Other lines are calculated for two closely space identical wires with $L_1=L_2=100$ nm and an empty gap of: $60$ nm (red short dashed line), $40$ nm (green long dashed line), $20$ nm (blue dash-dotted line), and $10$ nm (magenta dash-dot-dotted line).}
\end{center}
\end{figure}

The structure composed of two closely spaced wires shown in the inset of Fig. \ref{fig1} is examined first with an empty gap between the wires. Distinct surface plasmon-polariton (SPP) modes: longitudinal mode associated with oscillations of a charge density along the $z$-axis and transverse mode corresponding to oscillations in the direction perpendicular to $z$-axis \cite{doi:10.1021/la0513353}. 

Consider first the transmission characteristics of a single wire ($L_1=210$ nm). In the spectral regime displayed three peaks are shown. By examining their wire-length dependence and the corresponding charge densities \cite{doi:10.1021/la0513353} the lower frequency peak at $\omega = 0.98$ eV is identified as a longitudinal dipolar plasmon, the next one at $1.64$ eV is found to be a longitudinal quadrupole plasmon while the peak at $2.1$ eV corresponds to the transverse dipolar plasmon. For a single wire of length 100nm the longitudinal dipole mode peaks at $1.54$ eV, the longitudinal quadrupolar mode has moved to higher frequency above the range shown, while the transverse mode remains at $2.1$ eV. 

With a small gap separating the two wires one can expect to see a manifestation of the interaction between two longitudinal SPP modes supported by each wire. As two identical wires support modes with the same frequency, the close proximity of such modes permits the energy exchange between wires and thus lifts the degeneracy. This effect is clearly seen in Fig. \ref{fig1} as a splitting of the longitudinal SPP mode at $1.54$ eV. The splitting is  noticeable for a gap of $60$ nm since SPP modes are evanescent and decay exponentially with a distance from the surface of each wire, the gap becomes narrower the splitting significantly increases reaching $232$ meV for a $10$ nm gap. The observed Rabi splitting indicates the fact that interacting longitudinal SPP modes are in the so-called strong coupling regime permitting efficient energy exchange between wires \cite{0034-4885-78-1-013901}. By comparing transmission through a single wire (black solid line) and through two closely spaced wires with a gap of $10$ nm (magenta dash-dot-dotted line) we observe nearly perfect overlap of the quadrupole mode for a single wire with the high frequency mode in the system of coupled wires. Similar simulations comparing transmission through wider gaps with that through a single wire with a corresponding length confirm the quadrupole nature of the high frequency mode. It should be noted that the transmission in vacuum if wires are removed is several orders of magnitude smaller than that shown in Fig.\ref{fig1}.

\begin{figure}
\begin{center}
\includegraphics[width=0.9\textwidth]{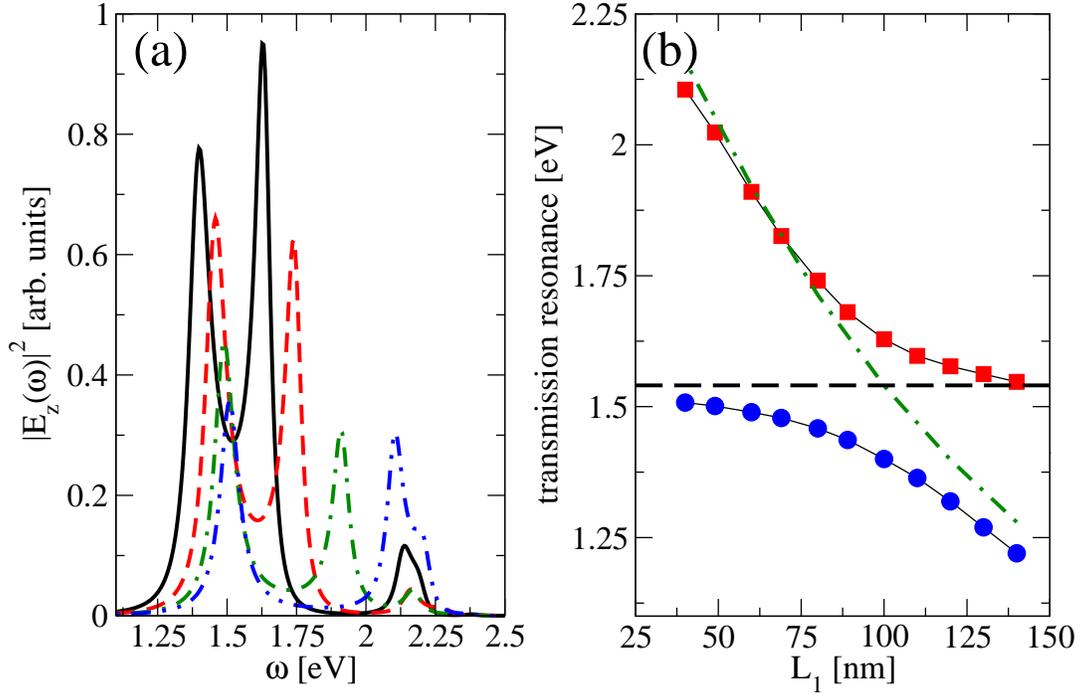}
\caption{\label{fig2} Panel (a) shows the transmitted intensity in the longitudinal direction, $\left | E_z(\omega)\right |^2$, as a function of the incident frequency, for the system of two wires with the fixed $L_2=100$ nm and variable $L_1$  and a gap of $\Delta L=10$ nm between the wires (see the inset of Fig. \ref{fig1} for details of the geometry). The black solid line shows results for symmetric system with $L_1=100$ nm, red dashed line shows data for $L_1=80$ nm, green dash-dotted line shows results for $L_1=60$ nm, and blue dash-dot-dotted line shows results for $L_1=40$ nm. Panel (b) shows the energies of two transmission resonances as functions of the wire's length $L_1$. Horizontal black dashed line indicates the energy of the longitudinal SPP resonance for a single $100$ nm wire. Green dash-dotted line shows how the energy of the longitudinal SPP resonance of a single wire varies with its length. Other parameters are the same as in Fig. \ref{fig1}.}
\end{center}
\end{figure}

To further scrutinize the strong coupling between interacting longitudinal SPP modes we perform a series of simulations varying the length of one wire while keeping the length of the other constant. This allows us to alter the frequency of one of the longitudinal SPP resonances sweeping through the other. The results of the simulations are shown in Fig. \ref{fig2}. We note that the high energy SPP mode seen in Fig. \ref{fig2}a at $2.16$ eV is the transverse SPP resonance position changes mildly from $2.15$ eV to $2.2$ eV when the length of one of the wires varies. The structure of the splitting of the longitudinal mode spectrum changes significantly, with the higher frequency peak changing much faster than the lower one. This apparent asymmetry is easily explained by referring to the avoided crossing behavior shown in Fig. \ref{fig2}b: when the length of the wire is comparable to its diameter it scatters light as a nanoparticle with a single dipolar Mie resonance of frequency similar to the transverse mode ($2.16$ eV). As the length increases this mode moves to the red while the unperturbed mode of the other wire of fixed length is unchanged. An avoided crossing is seen when the two lengths match.

\begin{figure}
\begin{center}
\includegraphics[width=0.9\textwidth]{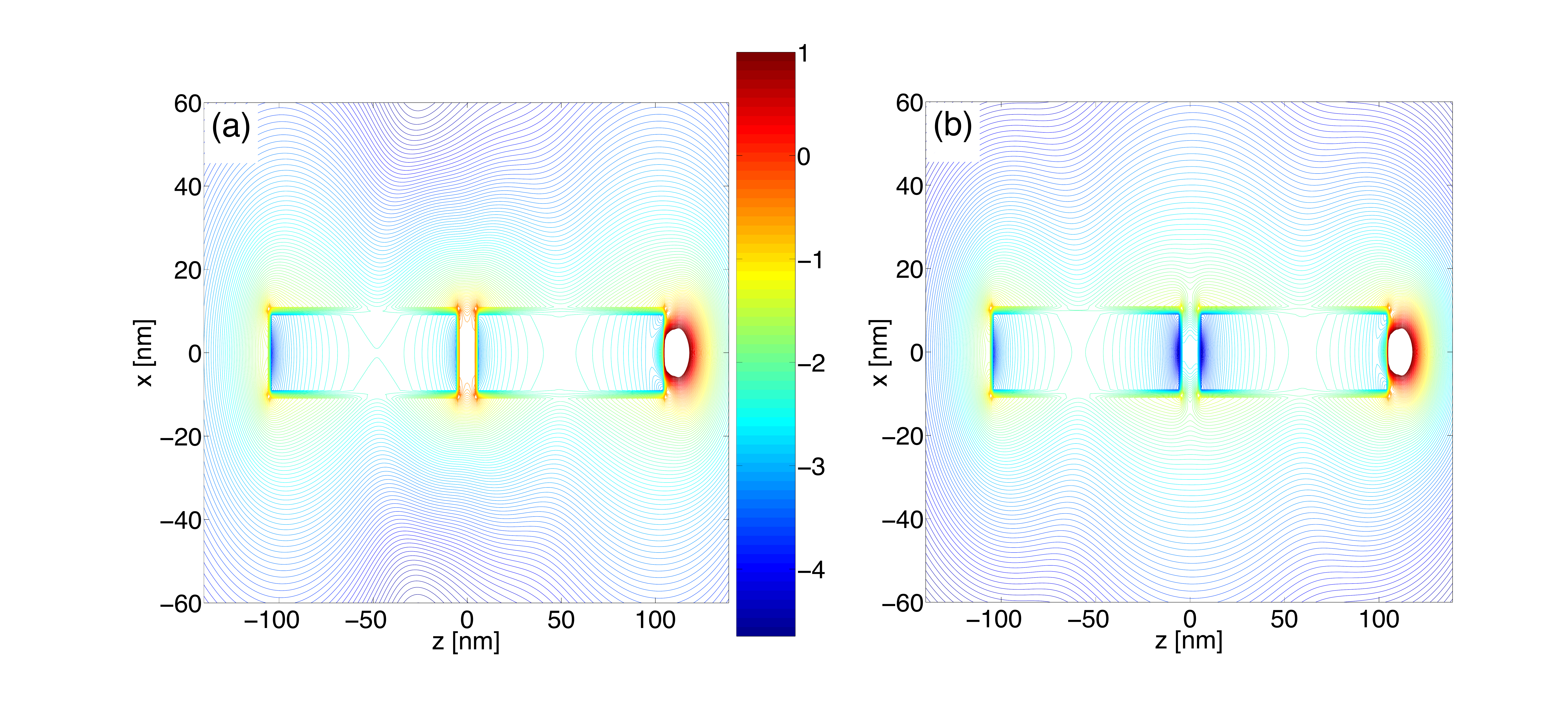}
\caption{\label{fig3} Steady-state electromagnetic intensity distributions for the system of two interacting wires separated by an empty gap of $10$ nm (see the inset in Fig. \ref{fig1} for details of the geometry). The length of each wire is $100$ nm. The system is excited by a pointwise dipole placed at the right side and oscillating in longitudinal $z$-direction. Panel (a) shows normalized electromagnetic intensity as a function of $x$ and $z$ (in nm) for the incident frequency of $1.40$ eV (the low energy resonance  for data shown as a magenta dash-dot-dotted line in Fig. (\ref{fig1})). Panel (b) shows normalized electromagnetic intensity calculated $1.63$ eV (the high energy resonance for data shown as a magenta dash-dot-dotted line in Fig. \ref{fig1}). The electromagnetic intensity distributions are plotted in logarithmic scale and normalized to the incident intensity.}
\end{center}
\end{figure}

Several other observations can be made. First, although the transversal modes of the two wires are in principle also coupled, this coupling is too weak to give a noticeable Rabi splitting. Second, even at the shortest length similar to the diameter, the plasmon frequency associated with the short wire is affected by the presence of the other wire and slightly deviate from the value obtained for the isolated wire (green dash-dotted line in Fig.  \ref{fig2}b). This probably results from the damping and frequency shift affected by the proximity to a dissipative dielectric object (the second wire) but may also reflect the effect of interaction with the transversal plasmon of this second wire. As expected, similar results were obtained when we varied $L_2$ with $L_1$ being kept constant.

Finally we note that these observations may be sensitive to the polarization of the injecting source. We defer the study of this issue to later work. 

In order to understand the physics behind the Rabi splitting we examine spatial distributions of the electromagnetic energy at two resonant frequencies. In the strong coupling regime the system of two wires forms two states as illustrated in Fig. \ref{fig3}. The low energy mode has a maximum in the gap while the high energy mode has a node. Obviously we are seeing the equivalent of the behavior of bonding and antibonding orbitals in a system of two coupled emitters \cite{Halas:2011aa} as this was confirmed in Fig. \ref{fig1} by comparing transmission through a single wire with that through two closely spaced wires.

\begin{figure}[t!]
\begin{center}
\includegraphics[width=0.9\textwidth]{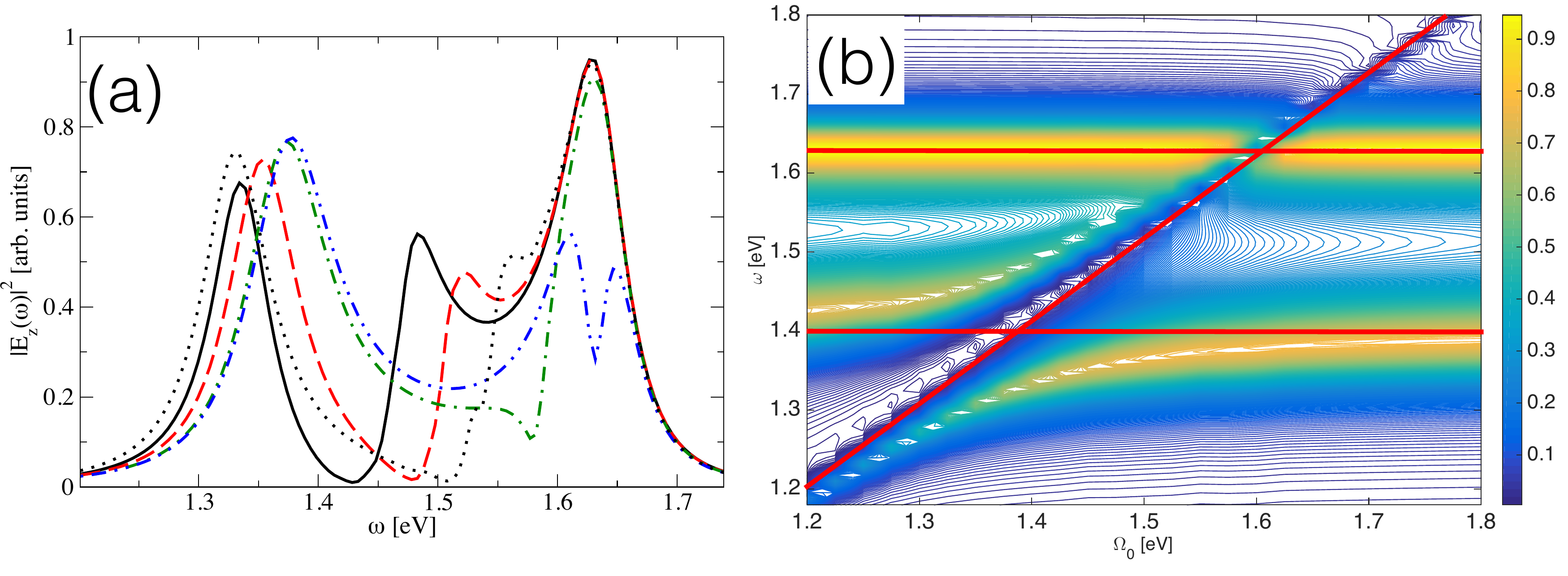}
\caption{\label{fig4} Coupled plasmon-exciton system. Panel (a) shows transmission as a function of the incident frequency calculated for two wires with a length of $L_1=L_2=100$ nm with a molecular aggregate placed in the gap between the wires ($\Delta L=10$ wide). The molecular transition frequency $\Omega_0=1.40$ eV corresponds to the solid and dotted black lines (see below), dashed red line shows data for $\Omega_0=1.45$ eV, dash-dotted green line shows results for $\Omega_0=1.55$ eV, and dash-dot-dotted blue line shows data for $\Omega_0=1.60$ eV. The molecular number density is $5\times10^{25}$ m$^{-3}$ for all lines except for the dotted black line, which is calculated at $10^{26}$ m$^{-3}$. Panel (b) shows transmission as a function of the incident frequency and molecular transition frequency, $\Omega_0$. Two horizontal red lines show the energies of the hybrid SP modes without molecules in the gap. The third red line with a slope shows the molecular transition frequency sweep.}
\end{center}
\end{figure}

Next we add a molecular aggregate filling the gap between the interacting wires and investigate how molecules resonant to longitudinal SPP modes modify the transmission spectrum. The values of the molecular parameters used in this work are: the molecular transition dipole $\mu_{12}=25$ Debye, $\gamma_{21}=6.892\times10^{-4}$ eV (corresponding to $6$ ps), $\gamma_{d}=6.565\times10^{-3}$ eV (corresponding to $630$ fs). When a molecular aggregate is placed inside the gap the local electromagnetic field is coupled to molecules. This system is analogous to three coupled oscillators: two longitudinal SPP electromagnetic modes and molecular excitons. The ratio of intensities spatially averaged over gap's volume for the low energy mode to the high one for the parameters in Fig. \ref{fig3} is $59.95$. The coupling strength depends on a local field amplitude. One can anticipate that the low energy plasmon mode with a higher amplitude in the gap would couple to molecules appreciably stronger than the mode at the higher energy. As noted above, the antibonding mode is close in frequency to the quadrupolar mode of a single wire of similar total length.

Fig. \ref{fig4} shows results of simulations with varying molecular transition energy, $\Omega_0$. With molecules having a transition close to the low energy mode the transmission exhibits three resonances with actual transmission dropping by more than $2$ orders of magnitude for the incident photon energy slightly higher than the molecular transition (black solid line, Fig. \ref{fig4}a, $\omega_{\text{inc}}=1.432$ eV). Additional simulations (not shown) were carried out to calculate a spatial distribution of the electromagnetic energy for parameters corresponding to very low transmission. We found that in this case the molecular aggregate acts as a very efficient absorber dropping the overall transmission. When the molecular transition energy is around $1.54$ eV the transmission exhibits signs of interference effects as its shape has a clear Fano lineshape (green dash-dotted line, Fig. \ref{fig4}a). However at energies close to the high energy mode near $1.62$ eV the transmission looks nearly unperturbed compared to the case without molecules. This can be explained if we recall the fact that the high energy mode has a node in the gap making the coupling with molecules negligibly small.

Fig. \ref{fig4}b shows results of the fine sweep over the molecular transition frequency, $\Omega_0$. Since the molecular aggregate is placed in the gap between two wires one would expect to observe coupling between the low frequency SP mode (mainly localized in the gap) and nearly no effect on the antibonding mode. Such a regime is manifested by a clear avoided crossing near $1.4$ eV with the Rabi splitting reaching about $150$ meV. The effect of resonant molecules on the antibonding mode located near $1.63$ eV is not trivial. Although there is no strong coupling observed molecules act as absorbers resulting in a narrow minimum occurring inside the transmission maximum (see blue dash-dot-dotted line in Fig. \ref{fig4}a).

\begin{figure}[t!]
\begin{center}
\includegraphics[width=0.9\textwidth]{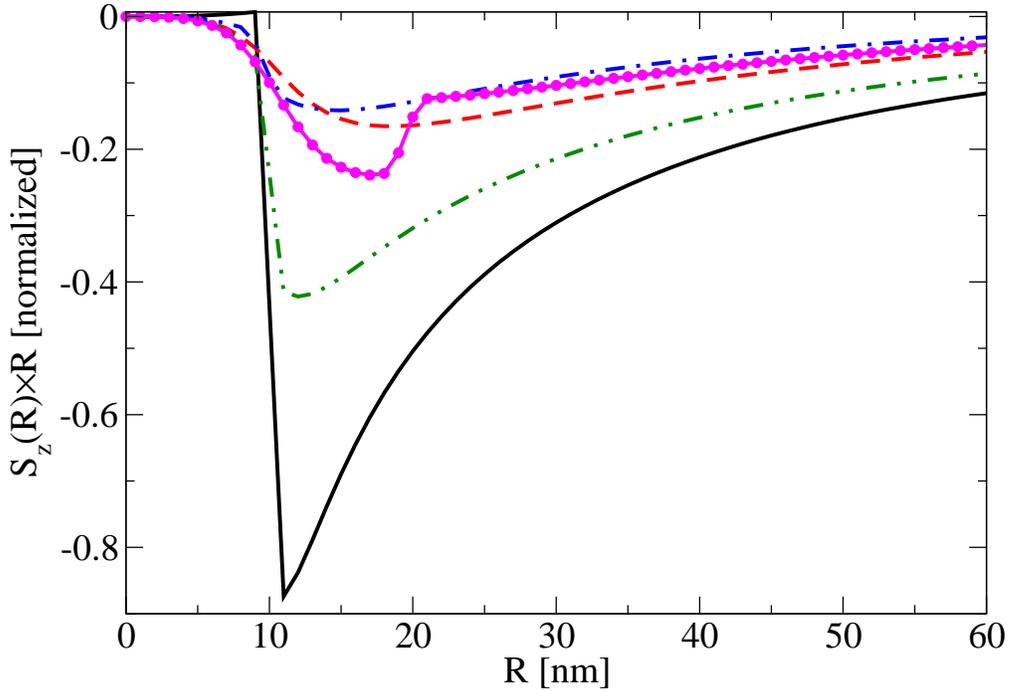}
\caption{\label{fig5} The $z$-component of the Poynting vector multiplied by the distance from the axis connecting wires, $S_z\times R$, as a function of $R$. $S_z$ is calculated at the center of the system at $z=0$. Black solid line shows results for a single wire $210$ nm long calculated at $\omega = 0.984$ eV (see black line in Fig. \ref{fig1}). Red dashed line presents results for two wires with $L_1=L_2=100$ nm and an empty gap of $\Delta L=10$ nm calculated at $\omega=1.4$ eV (see magenta dash-dot-dotted line in Fig. \ref{fig1}). Green dash-dot-dotted line shows results for two wires with the same characteristics as for the red line but with a gap filled with a non-dispersive dielectric with $\varepsilon=5$ calculated at $\omega=1.24$ eV (corresponding to the maximum transmission). Blue dash-dotted line shows results for the same system of two wires with the gap filled with molecules. The latter is calculated at $\omega=1.33$ eV (see black solid line in Fig. \ref{fig4}a). Magenta line connecting circles shows results for two wires with $L_1=L_2=100$ nm and gap of $\Delta L=10$ nm submerged in molecular cylinder of the radius $20$ nm. It is calculated at $\omega=1.58$ eV. Molecules for all calculations are resonant at $\Omega_0=1.4$ eV and the number density is $n_0=5\times 10^{25}$ m$^{-3}$. All wires are $20$ nm in diameter.}
\end{center}
\end{figure}

To investigate how electromagnetic energy is transported through a system of closely spaced nanowires we performed series of simulations varying the physical environment of the gap. First set of simulations was carried out with non-resonant molecules filling the gap. We found that there is almost no effect on transmission if the molecular transition energy is below $1.2$ eV irrespective of how high the molecular concentration is. This suggested another test, namely to replace molecules with a perfect electric conductor (PEC) completely closing the gap between the wires. Simulations showed that the effect of PEC on transmission is a red-shift of the low energy mode while the energy of the "antibonding" state remains constant. For the system with $\Delta L=10$ nm and $L_1=L_2=100$ nm the shift of the low energy mode is $0.4$ eV. These findings indicate that the electromagnetic energy transport occurs primarily along the surface of wires. Additional simulations examining spatial distributions of the Poynting vector confirmed that hypothesis as illustrated in Fig. \ref{fig5}. The $z$-component of the Poynting vector is evaluated for several representative cases as a function of the radial distance from the symmetry axis connecting two wires. We note that the negative sign of the energy flux corresponds to the propagation of EM energy from the source towards the detector (see the inset in Fig. \ref{fig1}). The magenta line shows simulations performed for a system comprised of two wires fully submerged in a cylinder composed of molecules. This system exhibit several resonances in the transmission spectrum (not shown). The "bonding" mode (see Fig. \ref{fig1}, blue dash-dotted line, resonance at $1.6$ eV, see also Fig. \ref{fig3}b) splits into two modes due to strong coupling between molecules and the plasmon mode. The frequency at which the energy flux is evaluated corresponds to the maximum transmission at $1.58$ eV. One can clearly see the EM energy mostly localized in the molecular layer $10$ nm $<R<20$ nm.
All five cases clearly demonstrate the fact that the propagation mainly occurs along the surface of wires.

Finally, the following observation from our simulation may appear surprising: the result shown in Fig. \ref{fig4} are not sensitive to the nature of the relaxation processes modeled in Eqs. (\ref{rate_equations}) in the following sense: If the magnitudes of the population and dephasing relaxation rates are varied such that $\gamma_{21}+2\gamma_{d} = \text{constant}$ the same spectrum is produced. This behavior characterizes systems in which spontaneous emission does not play a significant part in the spectral response (in the present calculation using classical EM field it is simply ignored). In such systems the optical response is dominated by the classical molecular polarization which is sensitive only to the sum $\gamma_{21}+2\gamma_{d}$.

\section{Conclusion}
\label{conclusion}
Using a numerical scheme based on the coupled Maxwell-Bloch equations implemented within the FDTD electromagnetic numerical solver, we have studied the transfer of electromagnetic radiation across a molecular gap separating two co-axial metal cylinders. In the absence of the molecular spacer in the gap separating the rods this system exhibits the consequence of plasmon-plasmon coupling, in particular between longitudinal plasmons in the two rods. The nature and magnitude of this coupling can be studied by varying the gap width as well as by changing the length of one wire keeping the other fixed, which produces the familiar non-crossing behavior. As a function of frequency the transmission is dominated by a splitted longitudinal plasmon peak. The two hybrid modes are the dipole-like "bonding" mode characterized by a peak intensity in the gap, and a quadrupole-like "antibonding" mode whose amplitude vanishes at the gap center.

When off resonant $2-$level molecules are placed in the gap, almost no effect on the frequency dependent transmission is observed. We have traced this behavior to the observation that much of the transmission takes place along the cylinders' edge. The "bonding" mode is significantly affected by the dielectric properties of the gap. In contrast, when the molecular system is in resonance with the plasmonic lineshape, the transmission is strongly modified, showing characteristics of strong exciton-plasmon coupling, modifying mostly the transmission near the lower frequency "bonding" plasmon mode. It is interesting to note that the presence resonant molecules species in the gap affects not only the molecule-field interaction but also the spatial distribution of the field intensity and the electromagnetic energy flux across the junction.

This study can be extended in several ways. First and obvious is the fact that the observation reported here can be sensitive to the molecular orientation in the gap. In the present study we have assumed that this orientation is random, implying an average isotropic molecular response, and a preferred molecular orientation may be an important parameter in determining the transmission across the junction. A more subtle issue is the role played by molecular fluorescence. In present experimental studies the fluorescence signal from dye molecules placed along the nano waveguide is used to report on the electromagnetic field distribution along the guide \cite{Paul:2012aa,Solis:2013aa}. In the configuration used in the present study spontaneous emission by strongly fluorescent molecule may play an active role in the observed optical response. To study such one needs to go beyond the current level of description that treat the electromagnetic field completely classically. We defer such considerations to future work.

\section*{Acknowledgements}
Both authors acknowledge support by the collaborative BSF grant No. 2014113. M.S. is also grateful for financial support by AFOSR grant No. FA9550-15-1-0189. The research of A.N. is also supported by the Israel Science Foundation.


\begin{thebibliography}{93}%
\makeatletter
\providecommand \@ifxundefined [1]{%
 \@ifx{#1\undefined}
}%
\providecommand \@ifnum [1]{%
 \ifnum #1\expandafter \@firstoftwo
 \else \expandafter \@secondoftwo
 \fi
}%
\providecommand \@ifx [1]{%
 \ifx #1\expandafter \@firstoftwo
 \else \expandafter \@secondoftwo
 \fi
}%
\providecommand \natexlab [1]{#1}%
\providecommand \enquote  [1]{``#1''}%
\providecommand \bibnamefont  [1]{#1}%
\providecommand \bibfnamefont [1]{#1}%
\providecommand \citenamefont [1]{#1}%
\providecommand \href@noop [0]{\@secondoftwo}%
\providecommand \href [0]{\begingroup \@sanitize@url \@href}%
\providecommand \@href[1]{\@@startlink{#1}\@@href}%
\providecommand \@@href[1]{\endgroup#1\@@endlink}%
\providecommand \@sanitize@url [0]{\catcode `\\12\catcode `\$12\catcode
  `\&12\catcode `\#12\catcode `\^12\catcode `\_12\catcode `\%12\relax}%
\providecommand \@@startlink[1]{}%
\providecommand \@@endlink[0]{}%
\providecommand \url  [0]{\begingroup\@sanitize@url \@url }%
\providecommand \@url [1]{\endgroup\@href {#1}{\urlprefix }}%
\providecommand \urlprefix  [0]{URL }%
\providecommand \Eprint [0]{\href }%
\providecommand \doibase [0]{http://dx.doi.org/}%
\providecommand \selectlanguage [0]{\@gobble}%
\providecommand \bibinfo  [0]{\@secondoftwo}%
\providecommand \bibfield  [0]{\@secondoftwo}%
\providecommand \translation [1]{[#1]}%
\providecommand \BibitemOpen [0]{}%
\providecommand \bibitemStop [0]{}%
\providecommand \bibitemNoStop [0]{.\EOS\space}%
\providecommand \EOS [0]{\spacefactor3000\relax}%
\providecommand \BibitemShut  [1]{\csname bibitem#1\endcsname}%
\let\auto@bib@innerbib\@empty
\bibitem [{\citenamefont {Maradudin}\ \emph {et~al.}(2014)\citenamefont
  {Maradudin}, \citenamefont {Sambles},\ and\ \citenamefont
  {Barnes}}]{maradudin2014modern}%
  \BibitemOpen
  \bibfield  {author} {\bibinfo {author} {\bibfnamefont {A.~A.}\ \bibnamefont
  {Maradudin}}, \bibinfo {author} {\bibfnamefont {J.~R.}\ \bibnamefont
  {Sambles}}, \ and\ \bibinfo {author} {\bibfnamefont {W.~L.}\ \bibnamefont
  {Barnes}},\ }\href@noop {} {\emph {\bibinfo {title} {Modern Plasmonics}}},\
  Vol.~\bibinfo {volume} {4}\ (\bibinfo  {publisher} {Elsevier},\ \bibinfo
  {year} {2014})\BibitemShut {NoStop}%
\bibitem [{\citenamefont {Nikoobakht}\ and\ \citenamefont
  {El-Sayed}(2003)}]{Nikoobakht:2003aa}%
  \BibitemOpen
  \bibfield  {author} {\bibinfo {author} {\bibfnamefont {B.}~\bibnamefont
  {Nikoobakht}}\ and\ \bibinfo {author} {\bibfnamefont {M.~A.}\ \bibnamefont
  {El-Sayed}},\ }\href {\doibase 10.1021/jp026770+} {\bibfield  {journal}
  {\bibinfo  {journal} {J. Chem. Phys. A}\ }\textbf {\bibinfo {volume} {107}},\
  \bibinfo {pages} {3372} (\bibinfo {year} {2003})}\BibitemShut {NoStop}%
\bibitem [{\citenamefont {Gantzounis}\ \emph {et~al.}(2005)\citenamefont
  {Gantzounis}, \citenamefont {Stefanou},\ and\ \citenamefont
  {Yannopapas}}]{Gantzounis:2005aa}%
  \BibitemOpen
  \bibfield  {author} {\bibinfo {author} {\bibfnamefont {G.}~\bibnamefont
  {Gantzounis}}, \bibinfo {author} {\bibfnamefont {N.}~\bibnamefont
  {Stefanou}}, \ and\ \bibinfo {author} {\bibfnamefont {V.}~\bibnamefont
  {Yannopapas}},\ }\href {http://stacks.iop.org/0953-8984/17/i=12/a=003}
  {\bibfield  {journal} {\bibinfo  {journal} {J. Phys. - Condens. Mat.}\
  }\textbf {\bibinfo {volume} {17}},\ \bibinfo {pages} {1791} (\bibinfo {year}
  {2005})}\BibitemShut {NoStop}%
\bibitem [{\citenamefont {Talley}\ \emph {et~al.}(2005)\citenamefont {Talley},
  \citenamefont {Jackson}, \citenamefont {Oubre}, \citenamefont {Grady},
  \citenamefont {Hollars}, \citenamefont {Lane}, \citenamefont {Huser},
  \citenamefont {Nordlander},\ and\ \citenamefont {Halas}}]{Talley:2005aa}%
  \BibitemOpen
  \bibfield  {author} {\bibinfo {author} {\bibfnamefont {C.~E.}\ \bibnamefont
  {Talley}}, \bibinfo {author} {\bibfnamefont {J.~B.}\ \bibnamefont {Jackson}},
  \bibinfo {author} {\bibfnamefont {C.}~\bibnamefont {Oubre}}, \bibinfo
  {author} {\bibfnamefont {N.~K.}\ \bibnamefont {Grady}}, \bibinfo {author}
  {\bibfnamefont {C.~W.}\ \bibnamefont {Hollars}}, \bibinfo {author}
  {\bibfnamefont {S.~M.}\ \bibnamefont {Lane}}, \bibinfo {author}
  {\bibfnamefont {T.~R.}\ \bibnamefont {Huser}}, \bibinfo {author}
  {\bibfnamefont {P.}~\bibnamefont {Nordlander}}, \ and\ \bibinfo {author}
  {\bibfnamefont {N.~J.}\ \bibnamefont {Halas}},\ }\href {\doibase
  10.1021/nl050928v} {\bibfield  {journal} {\bibinfo  {journal} {Nano Lett.}\
  }\textbf {\bibinfo {volume} {5}},\ \bibinfo {pages} {1569} (\bibinfo {year}
  {2005})}\BibitemShut {NoStop}%
\bibitem [{\citenamefont {Imura}\ \emph {et~al.}(2006)\citenamefont {Imura},
  \citenamefont {Okamoto}, \citenamefont {Hossain},\ and\ \citenamefont
  {Kitajima}}]{imura2006near}%
  \BibitemOpen
  \bibfield  {author} {\bibinfo {author} {\bibfnamefont {K.}~\bibnamefont
  {Imura}}, \bibinfo {author} {\bibfnamefont {H.}~\bibnamefont {Okamoto}},
  \bibinfo {author} {\bibfnamefont {M.~K.}\ \bibnamefont {Hossain}}, \ and\
  \bibinfo {author} {\bibfnamefont {M.}~\bibnamefont {Kitajima}},\ }\href@noop
  {} {\bibfield  {journal} {\bibinfo  {journal} {Chem. Lett.}\ }\textbf
  {\bibinfo {volume} {35}},\ \bibinfo {pages} {78} (\bibinfo {year}
  {2006})}\BibitemShut {NoStop}%
\bibitem [{\citenamefont {Jain}\ \emph {et~al.}(2006)\citenamefont {Jain},
  \citenamefont {Eustis},\ and\ \citenamefont {El-Sayed}}]{Jain:2006aa}%
  \BibitemOpen
  \bibfield  {author} {\bibinfo {author} {\bibfnamefont {P.~K.}\ \bibnamefont
  {Jain}}, \bibinfo {author} {\bibfnamefont {S.}~\bibnamefont {Eustis}}, \ and\
  \bibinfo {author} {\bibfnamefont {M.~A.}\ \bibnamefont {El-Sayed}},\ }\href
  {\doibase 10.1021/jp063879z} {\bibfield  {journal} {\bibinfo  {journal} {J.
  Phys. Chem. B}\ }\textbf {\bibinfo {volume} {110}},\ \bibinfo {pages} {18243}
  (\bibinfo {year} {2006})}\BibitemShut {NoStop}%
\bibitem [{\citenamefont {Jain}\ \emph {et~al.}(2007)\citenamefont {Jain},
  \citenamefont {Huang},\ and\ \citenamefont {El-Sayed}}]{Jain:2007aa}%
  \BibitemOpen
  \bibfield  {author} {\bibinfo {author} {\bibfnamefont {P.~K.}\ \bibnamefont
  {Jain}}, \bibinfo {author} {\bibfnamefont {W.}~\bibnamefont {Huang}}, \ and\
  \bibinfo {author} {\bibfnamefont {M.~A.}\ \bibnamefont {El-Sayed}},\ }\href
  {\doibase 10.1021/nl071008a} {\bibfield  {journal} {\bibinfo  {journal} {Nano
  Lett.}\ }\textbf {\bibinfo {volume} {7}},\ \bibinfo {pages} {2080} (\bibinfo
  {year} {2007})}\BibitemShut {NoStop}%
\bibitem [{\citenamefont {Yannopapas}\ and\ \citenamefont
  {Vitanov}(2007)}]{0953-8984-19-9-096210}%
  \BibitemOpen
  \bibfield  {author} {\bibinfo {author} {\bibfnamefont {V.}~\bibnamefont
  {Yannopapas}}\ and\ \bibinfo {author} {\bibfnamefont {N.~V.}\ \bibnamefont
  {Vitanov}},\ }\href {http://stacks.iop.org/0953-8984/19/i=9/a=096210}
  {\bibfield  {journal} {\bibinfo  {journal} {J. Phys. - Condens. Mat.}\
  }\textbf {\bibinfo {volume} {19}},\ \bibinfo {pages} {096210} (\bibinfo
  {year} {2007})}\BibitemShut {NoStop}%
\bibitem [{\citenamefont {Jain}\ and\ \citenamefont
  {El-Sayed}(2008)}]{Jain:2008aa}%
  \BibitemOpen
  \bibfield  {author} {\bibinfo {author} {\bibfnamefont {P.~K.}\ \bibnamefont
  {Jain}}\ and\ \bibinfo {author} {\bibfnamefont {M.~A.}\ \bibnamefont
  {El-Sayed}},\ }\href {\doibase 10.1021/jp7120356} {\bibfield  {journal}
  {\bibinfo  {journal} {J. Phys. Chem. C}\ }\textbf {\bibinfo {volume} {112}},\
  \bibinfo {pages} {4954} (\bibinfo {year} {2008})}\BibitemShut {NoStop}%
\bibitem [{\citenamefont {Li}\ \emph {et~al.}(2009)\citenamefont {Li},
  \citenamefont {Camargo}, \citenamefont {Lu},\ and\ \citenamefont
  {Xia}}]{Li:2009aa}%
  \BibitemOpen
  \bibfield  {author} {\bibinfo {author} {\bibfnamefont {W.}~\bibnamefont
  {Li}}, \bibinfo {author} {\bibfnamefont {P.~H.~C.}\ \bibnamefont {Camargo}},
  \bibinfo {author} {\bibfnamefont {X.}~\bibnamefont {Lu}}, \ and\ \bibinfo
  {author} {\bibfnamefont {Y.}~\bibnamefont {Xia}},\ }\href {\doibase
  10.1021/nl803621x} {\bibfield  {journal} {\bibinfo  {journal} {Nano Lett.}\
  }\textbf {\bibinfo {volume} {9}},\ \bibinfo {pages} {485} (\bibinfo {year}
  {2009})}\BibitemShut {NoStop}%
\bibitem [{\citenamefont {Chergui}\ \emph {et~al.}(2009)\citenamefont
  {Chergui}, \citenamefont {Melikyan},\ and\ \citenamefont
  {Minassian}}]{Chergui:2009aa}%
  \BibitemOpen
  \bibfield  {author} {\bibinfo {author} {\bibfnamefont {M.}~\bibnamefont
  {Chergui}}, \bibinfo {author} {\bibfnamefont {A.}~\bibnamefont {Melikyan}}, \
  and\ \bibinfo {author} {\bibfnamefont {H.}~\bibnamefont {Minassian}},\ }\href
  {\doibase 10.1021/jp810646m} {\bibfield  {journal} {\bibinfo  {journal} {J.
  Chem. Phys. C}\ }\textbf {\bibinfo {volume} {113}},\ \bibinfo {pages} {6463}
  (\bibinfo {year} {2009})}\BibitemShut {NoStop}%
\bibitem [{\citenamefont {Miljkovi{\'c}}\ \emph {et~al.}(2010)\citenamefont
  {Miljkovi{\'c}}, \citenamefont {Pakizeh}, \citenamefont {Sepulveda},
  \citenamefont {Johansson},\ and\ \citenamefont
  {K{\"a}ll}}]{Miljkovic:2010aa}%
  \BibitemOpen
  \bibfield  {author} {\bibinfo {author} {\bibfnamefont {V.~D.}\ \bibnamefont
  {Miljkovi{\'c}}}, \bibinfo {author} {\bibfnamefont {T.}~\bibnamefont
  {Pakizeh}}, \bibinfo {author} {\bibfnamefont {B.}~\bibnamefont {Sepulveda}},
  \bibinfo {author} {\bibfnamefont {P.}~\bibnamefont {Johansson}}, \ and\
  \bibinfo {author} {\bibfnamefont {M.}~\bibnamefont {K{\"a}ll}},\ }\href
  {\doibase 10.1021/jp911371r} {\bibfield  {journal} {\bibinfo  {journal} {J.
  Chem. Phys. C}\ }\textbf {\bibinfo {volume} {114}},\ \bibinfo {pages} {7472}
  (\bibinfo {year} {2010})}\BibitemShut {NoStop}%
\bibitem [{\citenamefont {Wustholz}\ \emph {et~al.}(2010)\citenamefont
  {Wustholz}, \citenamefont {Henry}, \citenamefont {McMahon}, \citenamefont
  {Freeman}, \citenamefont {Valley}, \citenamefont {Piotti}, \citenamefont
  {Natan}, \citenamefont {Schatz},\ and\ \citenamefont
  {Duyne}}]{Wustholz:2010aa}%
  \BibitemOpen
  \bibfield  {author} {\bibinfo {author} {\bibfnamefont {K.~L.}\ \bibnamefont
  {Wustholz}}, \bibinfo {author} {\bibfnamefont {A.-I.}\ \bibnamefont {Henry}},
  \bibinfo {author} {\bibfnamefont {J.~M.}\ \bibnamefont {McMahon}}, \bibinfo
  {author} {\bibfnamefont {R.~G.}\ \bibnamefont {Freeman}}, \bibinfo {author}
  {\bibfnamefont {N.}~\bibnamefont {Valley}}, \bibinfo {author} {\bibfnamefont
  {M.~E.}\ \bibnamefont {Piotti}}, \bibinfo {author} {\bibfnamefont {M.~J.}\
  \bibnamefont {Natan}}, \bibinfo {author} {\bibfnamefont {G.~C.}\ \bibnamefont
  {Schatz}}, \ and\ \bibinfo {author} {\bibfnamefont {R.~P.~V.}\ \bibnamefont
  {Duyne}},\ }\href {\doibase 10.1021/ja104174m} {\bibfield  {journal}
  {\bibinfo  {journal} {J. Am. Chem. Soc.}\ }\textbf {\bibinfo {volume}
  {132}},\ \bibinfo {pages} {10903} (\bibinfo {year} {2010})}\BibitemShut
  {NoStop}%
\bibitem [{\citenamefont {Farcau}\ and\ \citenamefont
  {Astilean}(2010)}]{Farcau:2010aa}%
  \BibitemOpen
  \bibfield  {author} {\bibinfo {author} {\bibfnamefont {C.}~\bibnamefont
  {Farcau}}\ and\ \bibinfo {author} {\bibfnamefont {S.}~\bibnamefont
  {Astilean}},\ }\href {\doibase 10.1021/jp100861w} {\bibfield  {journal}
  {\bibinfo  {journal} {J. Chem. Phys. C}\ }\textbf {\bibinfo {volume} {114}},\
  \bibinfo {pages} {11717} (\bibinfo {year} {2010})}\BibitemShut {NoStop}%
\bibitem [{\citenamefont {Letnes}\ \emph {et~al.}(2011)\citenamefont {Letnes},
  \citenamefont {Simonsen},\ and\ \citenamefont {Mills}}]{Letnes:2011aa}%
  \BibitemOpen
  \bibfield  {author} {\bibinfo {author} {\bibfnamefont {P.~A.}\ \bibnamefont
  {Letnes}}, \bibinfo {author} {\bibfnamefont {I.}~\bibnamefont {Simonsen}}, \
  and\ \bibinfo {author} {\bibfnamefont {D.~L.}\ \bibnamefont {Mills}},\ }\href
  {http://link.aps.org/doi/10.1103/PhysRevB.83.075426} {\bibfield  {journal}
  {\bibinfo  {journal} {Phys. Rev. B}\ }\textbf {\bibinfo {volume} {83}},\
  \bibinfo {pages} {075426} (\bibinfo {year} {2011})}\BibitemShut {NoStop}%
\bibitem [{\citenamefont {Encina}\ and\ \citenamefont
  {Coronado}(2011)}]{Encina:2011aa}%
  \BibitemOpen
  \bibfield  {author} {\bibinfo {author} {\bibfnamefont {E.~R.}\ \bibnamefont
  {Encina}}\ and\ \bibinfo {author} {\bibfnamefont {E.~A.}\ \bibnamefont
  {Coronado}},\ }\href {\doibase 10.1021/jp205158w} {\bibfield  {journal}
  {\bibinfo  {journal} {J. Chem. Phys. C}\ }\textbf {\bibinfo {volume} {115}},\
  \bibinfo {pages} {15908} (\bibinfo {year} {2011})}\BibitemShut {NoStop}%
\bibitem [{\citenamefont {Whitmore}\ \emph {et~al.}(2011)\citenamefont
  {Whitmore}, \citenamefont {El-Khoury}, \citenamefont {Fabris}, \citenamefont
  {Chu}, \citenamefont {Bazan}, \citenamefont {Potma},\ and\ \citenamefont
  {Apkarian}}]{Whitmore:2011aa}%
  \BibitemOpen
  \bibfield  {author} {\bibinfo {author} {\bibfnamefont {D.}~\bibnamefont
  {Whitmore}}, \bibinfo {author} {\bibfnamefont {P.~Z.}\ \bibnamefont
  {El-Khoury}}, \bibinfo {author} {\bibfnamefont {L.}~\bibnamefont {Fabris}},
  \bibinfo {author} {\bibfnamefont {P.}~\bibnamefont {Chu}}, \bibinfo {author}
  {\bibfnamefont {G.~C.}\ \bibnamefont {Bazan}}, \bibinfo {author}
  {\bibfnamefont {E.~O.}\ \bibnamefont {Potma}}, \ and\ \bibinfo {author}
  {\bibfnamefont {V.~A.}\ \bibnamefont {Apkarian}},\ }\href {\doibase
  10.1021/jp205055h} {\bibfield  {journal} {\bibinfo  {journal} {J. Chem. Phys.
  C}\ }\textbf {\bibinfo {volume} {115}},\ \bibinfo {pages} {15900} (\bibinfo
  {year} {2011})}\BibitemShut {NoStop}%
\bibitem [{\citenamefont {Jung}\ \emph {et~al.}(2011)\citenamefont {Jung},
  \citenamefont {Trolle}, \citenamefont {Pedersen},\ and\ \citenamefont
  {Pedersen}}]{Jung:2011aa}%
  \BibitemOpen
  \bibfield  {author} {\bibinfo {author} {\bibfnamefont {J.}~\bibnamefont
  {Jung}}, \bibinfo {author} {\bibfnamefont {M.~L.}\ \bibnamefont {Trolle}},
  \bibinfo {author} {\bibfnamefont {K.}~\bibnamefont {Pedersen}}, \ and\
  \bibinfo {author} {\bibfnamefont {T.~G.}\ \bibnamefont {Pedersen}},\ }\href
  {http://link.aps.org/doi/10.1103/PhysRevB.84.165447} {\bibfield  {journal}
  {\bibinfo  {journal} {Phys. Rev. B}\ }\textbf {\bibinfo {volume} {84}},\
  \bibinfo {pages} {165447} (\bibinfo {year} {2011})}\BibitemShut {NoStop}%
\bibitem [{\citenamefont {Banik}\ \emph
  {et~al.}(2012{\natexlab{a}})\citenamefont {Banik}, \citenamefont {Nag},
  \citenamefont {El-Khoury}, \citenamefont {Rodriguez~Perez}, \citenamefont
  {Guarrotxena}, \citenamefont {Bazan},\ and\ \citenamefont
  {Apkarian}}]{Banik:2012aa}%
  \BibitemOpen
  \bibfield  {author} {\bibinfo {author} {\bibfnamefont {M.}~\bibnamefont
  {Banik}}, \bibinfo {author} {\bibfnamefont {A.}~\bibnamefont {Nag}}, \bibinfo
  {author} {\bibfnamefont {P.~Z.}\ \bibnamefont {El-Khoury}}, \bibinfo {author}
  {\bibfnamefont {A.}~\bibnamefont {Rodriguez~Perez}}, \bibinfo {author}
  {\bibfnamefont {N.}~\bibnamefont {Guarrotxena}}, \bibinfo {author}
  {\bibfnamefont {G.~C.}\ \bibnamefont {Bazan}}, \ and\ \bibinfo {author}
  {\bibfnamefont {V.~A.}\ \bibnamefont {Apkarian}},\ }\href {\doibase
  10.1021/jp302013k} {\bibfield  {journal} {\bibinfo  {journal} {J. Chem. Phys.
  C}\ }\textbf {\bibinfo {volume} {116}},\ \bibinfo {pages} {10415} (\bibinfo
  {year} {2012}{\natexlab{a}})}\BibitemShut {NoStop}%
\bibitem [{\citenamefont {Chang}\ \emph {et~al.}(2012)\citenamefont {Chang},
  \citenamefont {Willingham}, \citenamefont {Slaughter}, \citenamefont
  {Dominguez-Medina}, \citenamefont {Swanglap},\ and\ \citenamefont
  {Link}}]{Chang:2012aa}%
  \BibitemOpen
  \bibfield  {author} {\bibinfo {author} {\bibfnamefont {W.-S.}\ \bibnamefont
  {Chang}}, \bibinfo {author} {\bibfnamefont {B.}~\bibnamefont {Willingham}},
  \bibinfo {author} {\bibfnamefont {L.~S.}\ \bibnamefont {Slaughter}}, \bibinfo
  {author} {\bibfnamefont {S.}~\bibnamefont {Dominguez-Medina}}, \bibinfo
  {author} {\bibfnamefont {P.}~\bibnamefont {Swanglap}}, \ and\ \bibinfo
  {author} {\bibfnamefont {S.}~\bibnamefont {Link}},\ }\href {\doibase
  10.1021/ar200337u} {\bibfield  {journal} {\bibinfo  {journal} {Accounts Chem.
  Res.}\ }\textbf {\bibinfo {volume} {45}},\ \bibinfo {pages} {1936} (\bibinfo
  {year} {2012})}\BibitemShut {NoStop}%
\bibitem [{\citenamefont {Banik}\ \emph
  {et~al.}(2012{\natexlab{b}})\citenamefont {Banik}, \citenamefont {El-Khoury},
  \citenamefont {Nag}, \citenamefont {Rodriguez-Perez}, \citenamefont
  {Guarrottxena}, \citenamefont {Bazan},\ and\ \citenamefont
  {Apkarian}}]{Banik:2012ab}%
  \BibitemOpen
  \bibfield  {author} {\bibinfo {author} {\bibfnamefont {M.}~\bibnamefont
  {Banik}}, \bibinfo {author} {\bibfnamefont {P.~Z.}\ \bibnamefont
  {El-Khoury}}, \bibinfo {author} {\bibfnamefont {A.}~\bibnamefont {Nag}},
  \bibinfo {author} {\bibfnamefont {A.}~\bibnamefont {Rodriguez-Perez}},
  \bibinfo {author} {\bibfnamefont {N.}~\bibnamefont {Guarrottxena}}, \bibinfo
  {author} {\bibfnamefont {G.~C.}\ \bibnamefont {Bazan}}, \ and\ \bibinfo
  {author} {\bibfnamefont {V.~A.}\ \bibnamefont {Apkarian}},\ }\href {\doibase
  10.1021/nn304277n} {\bibfield  {journal} {\bibinfo  {journal} {ACS Nano}\
  }\textbf {\bibinfo {volume} {6}},\ \bibinfo {pages} {10343} (\bibinfo {year}
  {2012}{\natexlab{b}})}\BibitemShut {NoStop}%
\bibitem [{\citenamefont {Fraire}\ \emph {et~al.}(2013)\citenamefont {Fraire},
  \citenamefont {P{\'e}rez},\ and\ \citenamefont {Coronado}}]{Fraire:2013aa}%
  \BibitemOpen
  \bibfield  {author} {\bibinfo {author} {\bibfnamefont {J.~C.}\ \bibnamefont
  {Fraire}}, \bibinfo {author} {\bibfnamefont {L.~A.}\ \bibnamefont
  {P{\'e}rez}}, \ and\ \bibinfo {author} {\bibfnamefont {E.~A.}\ \bibnamefont
  {Coronado}},\ }\href {\doibase 10.1021/jp3123709} {\bibfield  {journal}
  {\bibinfo  {journal} {J. Chem. Phys. C}\ }\textbf {\bibinfo {volume} {117}},\
  \bibinfo {pages} {23090} (\bibinfo {year} {2013})}\BibitemShut {NoStop}%
\bibitem [{\citenamefont {Prodan}\ \emph {et~al.}(2003)\citenamefont {Prodan},
  \citenamefont {Radloff}, \citenamefont {Halas},\ and\ \citenamefont
  {Nordlander}}]{Prodan419}%
  \BibitemOpen
  \bibfield  {author} {\bibinfo {author} {\bibfnamefont {E.}~\bibnamefont
  {Prodan}}, \bibinfo {author} {\bibfnamefont {C.}~\bibnamefont {Radloff}},
  \bibinfo {author} {\bibfnamefont {N.~J.}\ \bibnamefont {Halas}}, \ and\
  \bibinfo {author} {\bibfnamefont {P.}~\bibnamefont {Nordlander}},\ }\href
  {\doibase 10.1126/science.1089171} {\bibfield  {journal} {\bibinfo  {journal}
  {Science}\ }\textbf {\bibinfo {volume} {302}},\ \bibinfo {pages} {419}
  (\bibinfo {year} {2003})}\BibitemShut {NoStop}%
\bibitem [{\citenamefont {Nordlander}\ \emph {et~al.}(2004)\citenamefont
  {Nordlander}, \citenamefont {Oubre}, \citenamefont {Prodan}, \citenamefont
  {Li},\ and\ \citenamefont {Stockman}}]{Nordlander:2004aa}%
  \BibitemOpen
  \bibfield  {author} {\bibinfo {author} {\bibfnamefont {P.}~\bibnamefont
  {Nordlander}}, \bibinfo {author} {\bibfnamefont {C.}~\bibnamefont {Oubre}},
  \bibinfo {author} {\bibfnamefont {E.}~\bibnamefont {Prodan}}, \bibinfo
  {author} {\bibfnamefont {K.}~\bibnamefont {Li}}, \ and\ \bibinfo {author}
  {\bibfnamefont {M.~I.}\ \bibnamefont {Stockman}},\ }\href {\doibase
  10.1021/nl049681c} {\bibfield  {journal} {\bibinfo  {journal} {Nano Lett.}\
  }\textbf {\bibinfo {volume} {4}},\ \bibinfo {pages} {899} (\bibinfo {year}
  {2004})}\BibitemShut {NoStop}%
\bibitem [{\citenamefont {Nordlander}\ and\ \citenamefont
  {Prodan}(2004)}]{Nordlander:2004ab}%
  \BibitemOpen
  \bibfield  {author} {\bibinfo {author} {\bibfnamefont {P.}~\bibnamefont
  {Nordlander}}\ and\ \bibinfo {author} {\bibfnamefont {E.}~\bibnamefont
  {Prodan}},\ }\href {\doibase 10.1021/nl0486160} {\bibfield  {journal}
  {\bibinfo  {journal} {Nano Lett.}\ }\textbf {\bibinfo {volume} {4}},\
  \bibinfo {pages} {2209} (\bibinfo {year} {2004})}\BibitemShut {NoStop}%
\bibitem [{\citenamefont {Prodan}\ and\ \citenamefont
  {Nordlander}(2004)}]{Prodan:2004aa}%
  \BibitemOpen
  \bibfield  {author} {\bibinfo {author} {\bibfnamefont {E.}~\bibnamefont
  {Prodan}}\ and\ \bibinfo {author} {\bibfnamefont {P.}~\bibnamefont
  {Nordlander}},\ }\href
  {http://scitation.aip.org/content/aip/journal/jcp/120/11/10.1063/1.1647518}
  {\bibfield  {journal} {\bibinfo  {journal} {J. Chem. Phys.}\ }\textbf
  {\bibinfo {volume} {120}},\ \bibinfo {pages} {5444} (\bibinfo {year}
  {2004})}\BibitemShut {NoStop}%
\bibitem [{\citenamefont {Halas}\ \emph {et~al.}(2011)\citenamefont {Halas},
  \citenamefont {Lal}, \citenamefont {Chang}, \citenamefont {Link},\ and\
  \citenamefont {Nordlander}}]{Halas:2011aa}%
  \BibitemOpen
  \bibfield  {author} {\bibinfo {author} {\bibfnamefont {N.~J.}\ \bibnamefont
  {Halas}}, \bibinfo {author} {\bibfnamefont {S.}~\bibnamefont {Lal}}, \bibinfo
  {author} {\bibfnamefont {W.-S.}\ \bibnamefont {Chang}}, \bibinfo {author}
  {\bibfnamefont {S.}~\bibnamefont {Link}}, \ and\ \bibinfo {author}
  {\bibfnamefont {P.}~\bibnamefont {Nordlander}},\ }\href {\doibase
  10.1021/cr200061k} {\bibfield  {journal} {\bibinfo  {journal} {Chem. Rev.}\
  }\textbf {\bibinfo {volume} {111}},\ \bibinfo {pages} {3913} (\bibinfo {year}
  {2011})}\BibitemShut {NoStop}%
\bibitem [{\citenamefont {Esteban}\ \emph {et~al.}(2012)\citenamefont
  {Esteban}, \citenamefont {Borisov}, \citenamefont {Nordlander},\ and\
  \citenamefont {Aizpurua}}]{Esteban:2012aa}%
  \BibitemOpen
  \bibfield  {author} {\bibinfo {author} {\bibfnamefont {R.}~\bibnamefont
  {Esteban}}, \bibinfo {author} {\bibfnamefont {A.~G.}\ \bibnamefont
  {Borisov}}, \bibinfo {author} {\bibfnamefont {P.}~\bibnamefont {Nordlander}},
  \ and\ \bibinfo {author} {\bibfnamefont {J.}~\bibnamefont {Aizpurua}},\
  }\href {http://dx.doi.org/10.1038/ncomms1806} {\bibfield  {journal} {\bibinfo
   {journal} {Nat. Commun.}\ }\textbf {\bibinfo {volume} {3}},\ \bibinfo
  {pages} {825} (\bibinfo {year} {2012})}\BibitemShut {NoStop}%
\bibitem [{\citenamefont {Schlather}\ \emph {et~al.}(2013)\citenamefont
  {Schlather}, \citenamefont {Large}, \citenamefont {Urban}, \citenamefont
  {Nordlander},\ and\ \citenamefont {Halas}}]{doi:10.1021/nl4014887}%
  \BibitemOpen
  \bibfield  {author} {\bibinfo {author} {\bibfnamefont {A.~E.}\ \bibnamefont
  {Schlather}}, \bibinfo {author} {\bibfnamefont {N.}~\bibnamefont {Large}},
  \bibinfo {author} {\bibfnamefont {A.~S.}\ \bibnamefont {Urban}}, \bibinfo
  {author} {\bibfnamefont {P.}~\bibnamefont {Nordlander}}, \ and\ \bibinfo
  {author} {\bibfnamefont {N.~J.}\ \bibnamefont {Halas}},\ }\href {\doibase
  10.1021/nl4014887} {\bibfield  {journal} {\bibinfo  {journal} {Nano Lett.}\
  }\textbf {\bibinfo {volume} {13}},\ \bibinfo {pages} {3281} (\bibinfo {year}
  {2013})}\BibitemShut {NoStop}%
\bibitem [{\citenamefont {Harris}\ \emph {et~al.}(2009)\citenamefont {Harris},
  \citenamefont {Arnold}, \citenamefont {Blaber},\ and\ \citenamefont
  {Ford}}]{Harris:2009aa}%
  \BibitemOpen
  \bibfield  {author} {\bibinfo {author} {\bibfnamefont {N.}~\bibnamefont
  {Harris}}, \bibinfo {author} {\bibfnamefont {M.~D.}\ \bibnamefont {Arnold}},
  \bibinfo {author} {\bibfnamefont {M.~G.}\ \bibnamefont {Blaber}}, \ and\
  \bibinfo {author} {\bibfnamefont {M.~J.}\ \bibnamefont {Ford}},\ }\href
  {\doibase 10.1021/jp8083869} {\bibfield  {journal} {\bibinfo  {journal} {J.
  Chem. Phys. C}\ }\textbf {\bibinfo {volume} {113}},\ \bibinfo {pages} {2784}
  (\bibinfo {year} {2009})}\BibitemShut {NoStop}%
\bibitem [{Note1()}]{Note1}%
  \BibitemOpen
  \bibinfo {note} {Other important recurring themes such as induction of
  transient dielectric properties, lifetime of radiative and non-radiative
  processes, and formation and utilization of hot electrons and local hotspots
  are not related to the present study}\BibitemShut {NoStop}%
\bibitem [{\citenamefont {Gersten}\ and\ \citenamefont
  {Nitzan}(1984)}]{Gersten:1984aa}%
  \BibitemOpen
  \bibfield  {author} {\bibinfo {author} {\bibfnamefont {J.~I.}\ \bibnamefont
  {Gersten}}\ and\ \bibinfo {author} {\bibfnamefont {A.}~\bibnamefont
  {Nitzan}},\ }\href {\doibase http://dx.doi.org/10.1016/0009-2614(84)85300-2}
  {\bibfield  {journal} {\bibinfo  {journal} {Chem. Phys. Lett.}\ }\textbf
  {\bibinfo {volume} {104}},\ \bibinfo {pages} {31} (\bibinfo {year}
  {1984})}\BibitemShut {NoStop}%
\bibitem [{\citenamefont {Hua}\ \emph {et~al.}(1985)\citenamefont {Hua},
  \citenamefont {Gersten},\ and\ \citenamefont {Nitzan}}]{Hua:1985aa}%
  \BibitemOpen
  \bibfield  {author} {\bibinfo {author} {\bibfnamefont {X.~M.}\ \bibnamefont
  {Hua}}, \bibinfo {author} {\bibfnamefont {J.~I.}\ \bibnamefont {Gersten}}, \
  and\ \bibinfo {author} {\bibfnamefont {A.}~\bibnamefont {Nitzan}},\ }\href
  {http://scitation.aip.org/content/aip/journal/jcp/83/7/10.1063/1.449120}
  {\bibfield  {journal} {\bibinfo  {journal} {J. Chem. Phys.}\ }\textbf
  {\bibinfo {volume} {83}},\ \bibinfo {pages} {3650} (\bibinfo {year}
  {1985})}\BibitemShut {NoStop}%
\bibitem [{\citenamefont {Andrew}\ and\ \citenamefont
  {Barnes}(2004)}]{Andrew1002}%
  \BibitemOpen
  \bibfield  {author} {\bibinfo {author} {\bibfnamefont {P.}~\bibnamefont
  {Andrew}}\ and\ \bibinfo {author} {\bibfnamefont {W.~L.}\ \bibnamefont
  {Barnes}},\ }\href {\doibase 10.1126/science.1102992} {\bibfield  {journal}
  {\bibinfo  {journal} {Science}\ }\textbf {\bibinfo {volume} {306}},\ \bibinfo
  {pages} {1002} (\bibinfo {year} {2004})}\BibitemShut {NoStop}%
\bibitem [{\citenamefont {Gersten}(2007)}]{Gersten:2007aa}%
  \BibitemOpen
  \bibfield  {author} {\bibinfo {author} {\bibfnamefont {J.~I.}\ \bibnamefont
  {Gersten}},\ }\href {\doibase 10.1007/s11468-007-9028-9} {\bibfield
  {journal} {\bibinfo  {journal} {Plasmonics}\ }\textbf {\bibinfo {volume}
  {2}},\ \bibinfo {pages} {65} (\bibinfo {year} {2007})}\BibitemShut {NoStop}%
\bibitem [{\citenamefont {Zhang}\ \emph {et~al.}(2007)\citenamefont {Zhang},
  \citenamefont {Fu},\ and\ \citenamefont {Lakowicz}}]{Zhang:2007aa}%
  \BibitemOpen
  \bibfield  {author} {\bibinfo {author} {\bibfnamefont {J.}~\bibnamefont
  {Zhang}}, \bibinfo {author} {\bibfnamefont {Y.}~\bibnamefont {Fu}}, \ and\
  \bibinfo {author} {\bibfnamefont {J.~R.}\ \bibnamefont {Lakowicz}},\ }\href
  {\doibase 10.1021/jp062665e} {\bibfield  {journal} {\bibinfo  {journal} {J.
  Chem. Phys. C}\ }\textbf {\bibinfo {volume} {111}},\ \bibinfo {pages} {50}
  (\bibinfo {year} {2007})}\BibitemShut {NoStop}%
\bibitem [{\citenamefont {Reil}\ \emph {et~al.}(2008)\citenamefont {Reil},
  \citenamefont {Hohenester}, \citenamefont {Krenn},\ and\ \citenamefont
  {Leitner}}]{Reil:2008aa}%
  \BibitemOpen
  \bibfield  {author} {\bibinfo {author} {\bibfnamefont {F.}~\bibnamefont
  {Reil}}, \bibinfo {author} {\bibfnamefont {U.}~\bibnamefont {Hohenester}},
  \bibinfo {author} {\bibfnamefont {J.~R.}\ \bibnamefont {Krenn}}, \ and\
  \bibinfo {author} {\bibfnamefont {A.}~\bibnamefont {Leitner}},\ }\href
  {\doibase 10.1021/nl801480m} {\bibfield  {journal} {\bibinfo  {journal} {Nano
  Lett.}\ }\textbf {\bibinfo {volume} {8}},\ \bibinfo {pages} {4128} (\bibinfo
  {year} {2008})}\BibitemShut {NoStop}%
\bibitem [{\citenamefont {Marocico}\ and\ \citenamefont
  {Knoester}(2009)}]{Marocico:2009aa}%
  \BibitemOpen
  \bibfield  {author} {\bibinfo {author} {\bibfnamefont {C.~A.}\ \bibnamefont
  {Marocico}}\ and\ \bibinfo {author} {\bibfnamefont {J.}~\bibnamefont
  {Knoester}},\ }\href {http://link.aps.org/doi/10.1103/PhysRevA.79.053816}
  {\bibfield  {journal} {\bibinfo  {journal} {Phys. Rev. A}\ }\textbf {\bibinfo
  {volume} {79}},\ \bibinfo {pages} {053816} (\bibinfo {year}
  {2009})}\BibitemShut {NoStop}%
\bibitem [{\citenamefont {Saini}\ \emph {et~al.}(2009)\citenamefont {Saini},
  \citenamefont {Srinivas},\ and\ \citenamefont {Bagchi}}]{Saini:2009aa}%
  \BibitemOpen
  \bibfield  {author} {\bibinfo {author} {\bibfnamefont {S.}~\bibnamefont
  {Saini}}, \bibinfo {author} {\bibfnamefont {G.}~\bibnamefont {Srinivas}}, \
  and\ \bibinfo {author} {\bibfnamefont {B.}~\bibnamefont {Bagchi}},\ }\href
  {\doibase 10.1021/jp806536w} {\bibfield  {journal} {\bibinfo  {journal} {J.
  Chem. Phys. B}\ }\textbf {\bibinfo {volume} {113}},\ \bibinfo {pages} {1817}
  (\bibinfo {year} {2009})}\BibitemShut {NoStop}%
\bibitem [{\citenamefont {Chung}\ \emph {et~al.}(2010)\citenamefont {Chung},
  \citenamefont {Leung},\ and\ \citenamefont {Tsai}}]{Chung2010}%
  \BibitemOpen
  \bibfield  {author} {\bibinfo {author} {\bibfnamefont {H.~Y.}\ \bibnamefont
  {Chung}}, \bibinfo {author} {\bibfnamefont {P.~T.}\ \bibnamefont {Leung}}, \
  and\ \bibinfo {author} {\bibfnamefont {D.~P.}\ \bibnamefont {Tsai}},\ }\href
  {\doibase 10.1007/s11468-010-9151-x} {\bibfield  {journal} {\bibinfo
  {journal} {Plasmonics}\ }\textbf {\bibinfo {volume} {5}},\ \bibinfo {pages}
  {363} (\bibinfo {year} {2010})}\BibitemShut {NoStop}%
\bibitem [{\citenamefont {Mart{\'\i}n-Cano}\ \emph {et~al.}(2010)\citenamefont
  {Mart{\'\i}n-Cano}, \citenamefont {Mart{\'\i}n-Moreno}, \citenamefont
  {Garc{\'\i}a-Vidal},\ and\ \citenamefont {Moreno}}]{Martin-Cano:2010aa}%
  \BibitemOpen
  \bibfield  {author} {\bibinfo {author} {\bibfnamefont {D.}~\bibnamefont
  {Mart{\'\i}n-Cano}}, \bibinfo {author} {\bibfnamefont {L.}~\bibnamefont
  {Mart{\'\i}n-Moreno}}, \bibinfo {author} {\bibfnamefont {F.~J.}\ \bibnamefont
  {Garc{\'\i}a-Vidal}}, \ and\ \bibinfo {author} {\bibfnamefont
  {E.}~\bibnamefont {Moreno}},\ }\href {\doibase 10.1021/nl101876f} {\bibfield
  {journal} {\bibinfo  {journal} {Nano Lett.}\ }\textbf {\bibinfo {volume}
  {10}},\ \bibinfo {pages} {3129} (\bibinfo {year} {2010})}\BibitemShut
  {NoStop}%
\bibitem [{\citenamefont {Su}\ \emph {et~al.}(2010)\citenamefont {Su},
  \citenamefont {Zhang}, \citenamefont {Zhou}, \citenamefont {Peng},\ and\
  \citenamefont {Wang}}]{Su:10}%
  \BibitemOpen
  \bibfield  {author} {\bibinfo {author} {\bibfnamefont {X.-R.}\ \bibnamefont
  {Su}}, \bibinfo {author} {\bibfnamefont {W.}~\bibnamefont {Zhang}}, \bibinfo
  {author} {\bibfnamefont {L.}~\bibnamefont {Zhou}}, \bibinfo {author}
  {\bibfnamefont {X.-N.}\ \bibnamefont {Peng}}, \ and\ \bibinfo {author}
  {\bibfnamefont {Q.-Q.}\ \bibnamefont {Wang}},\ }\href {\doibase
  10.1364/OE.18.006516} {\bibfield  {journal} {\bibinfo  {journal} {Opt.
  Express}\ }\textbf {\bibinfo {volume} {18}},\ \bibinfo {pages} {6516}
  (\bibinfo {year} {2010})}\BibitemShut {NoStop}%
\bibitem [{\citenamefont {Faessler}\ \emph {et~al.}(2011)\citenamefont
  {Faessler}, \citenamefont {Hrelescu}, \citenamefont {Lutich}, \citenamefont
  {Osinkina}, \citenamefont {Mayilo}, \citenamefont {J{\"a}ckel},\ and\
  \citenamefont {Feldmann}}]{Faessler:2011aa}%
  \BibitemOpen
  \bibfield  {author} {\bibinfo {author} {\bibfnamefont {V.}~\bibnamefont
  {Faessler}}, \bibinfo {author} {\bibfnamefont {C.}~\bibnamefont {Hrelescu}},
  \bibinfo {author} {\bibfnamefont {A.~A.}\ \bibnamefont {Lutich}}, \bibinfo
  {author} {\bibfnamefont {L.}~\bibnamefont {Osinkina}}, \bibinfo {author}
  {\bibfnamefont {S.}~\bibnamefont {Mayilo}}, \bibinfo {author} {\bibfnamefont
  {F.}~\bibnamefont {J{\"a}ckel}}, \ and\ \bibinfo {author} {\bibfnamefont
  {J.}~\bibnamefont {Feldmann}},\ }\href {\doibase
  http://dx.doi.org/10.1016/j.cplett.2011.03.088} {\bibfield  {journal}
  {\bibinfo  {journal} {Chem. Phys. Lett.}\ }\textbf {\bibinfo {volume}
  {508}},\ \bibinfo {pages} {67} (\bibinfo {year} {2011})}\BibitemShut
  {NoStop}%
\bibitem [{\citenamefont {Marocico}\ and\ \citenamefont
  {Knoester}(2011)}]{Marocico:2011aa}%
  \BibitemOpen
  \bibfield  {author} {\bibinfo {author} {\bibfnamefont {C.~A.}\ \bibnamefont
  {Marocico}}\ and\ \bibinfo {author} {\bibfnamefont {J.}~\bibnamefont
  {Knoester}},\ }\href {http://link.aps.org/doi/10.1103/PhysRevA.84.053824}
  {\bibfield  {journal} {\bibinfo  {journal} {Phys. Rev. A}\ }\textbf {\bibinfo
  {volume} {84}},\ \bibinfo {pages} {053824} (\bibinfo {year}
  {2011})}\BibitemShut {NoStop}%
\bibitem [{\citenamefont {Pustovit}\ and\ \citenamefont
  {Shahbazyan}(2011)}]{Pustovit:2011aa}%
  \BibitemOpen
  \bibfield  {author} {\bibinfo {author} {\bibfnamefont {V.~N.}\ \bibnamefont
  {Pustovit}}\ and\ \bibinfo {author} {\bibfnamefont {T.~V.}\ \bibnamefont
  {Shahbazyan}},\ }\href {http://link.aps.org/doi/10.1103/PhysRevB.83.085427}
  {\bibfield  {journal} {\bibinfo  {journal} {Phys. Rev. B}\ }\textbf {\bibinfo
  {volume} {83}},\ \bibinfo {pages} {085427} (\bibinfo {year}
  {2011})}\BibitemShut {NoStop}%
\bibitem [{\citenamefont {Ant{\'o}n}\ \emph {et~al.}(2012)\citenamefont
  {Ant{\'o}n}, \citenamefont {Carre{\~n}o}, \citenamefont {Melle},
  \citenamefont {Calder{\'o}n}, \citenamefont {Cabrera-Granado}, \citenamefont
  {Cox},\ and\ \citenamefont {Singh}}]{Anton:2012aa}%
  \BibitemOpen
  \bibfield  {author} {\bibinfo {author} {\bibfnamefont {M.~A.}\ \bibnamefont
  {Ant{\'o}n}}, \bibinfo {author} {\bibfnamefont {F.}~\bibnamefont
  {Carre{\~n}o}}, \bibinfo {author} {\bibfnamefont {S.}~\bibnamefont {Melle}},
  \bibinfo {author} {\bibfnamefont {O.~G.}\ \bibnamefont {Calder{\'o}n}},
  \bibinfo {author} {\bibfnamefont {E.}~\bibnamefont {Cabrera-Granado}},
  \bibinfo {author} {\bibfnamefont {J.}~\bibnamefont {Cox}}, \ and\ \bibinfo
  {author} {\bibfnamefont {M.~R.}\ \bibnamefont {Singh}},\ }\href
  {http://link.aps.org/doi/10.1103/PhysRevB.86.155305} {\bibfield  {journal}
  {\bibinfo  {journal} {Phys. Rev. B}\ }\textbf {\bibinfo {volume} {86}},\
  \bibinfo {pages} {155305} (\bibinfo {year} {2012})}\BibitemShut {NoStop}%
\bibitem [{\citenamefont {Lunz}\ \emph {et~al.}(2012)\citenamefont {Lunz},
  \citenamefont {Zhang}, \citenamefont {Gerard}, \citenamefont {Gun'ko},
  \citenamefont {Lesnyak}, \citenamefont {Gaponik}, \citenamefont {Susha},
  \citenamefont {Rogach},\ and\ \citenamefont {Bradley}}]{Lunz:2012aa}%
  \BibitemOpen
  \bibfield  {author} {\bibinfo {author} {\bibfnamefont {M.}~\bibnamefont
  {Lunz}}, \bibinfo {author} {\bibfnamefont {X.}~\bibnamefont {Zhang}},
  \bibinfo {author} {\bibfnamefont {V.~A.}\ \bibnamefont {Gerard}}, \bibinfo
  {author} {\bibfnamefont {Y.~K.}\ \bibnamefont {Gun'ko}}, \bibinfo {author}
  {\bibfnamefont {V.}~\bibnamefont {Lesnyak}}, \bibinfo {author} {\bibfnamefont
  {N.}~\bibnamefont {Gaponik}}, \bibinfo {author} {\bibfnamefont {A.~S.}\
  \bibnamefont {Susha}}, \bibinfo {author} {\bibfnamefont {A.~L.}\ \bibnamefont
  {Rogach}}, \ and\ \bibinfo {author} {\bibfnamefont {A.~L.}\ \bibnamefont
  {Bradley}},\ }\href {\doibase 10.1021/jp309660s} {\bibfield  {journal}
  {\bibinfo  {journal} {J. Chem. Phys. C}\ }\textbf {\bibinfo {volume} {116}},\
  \bibinfo {pages} {26529} (\bibinfo {year} {2012})}\BibitemShut {NoStop}%
\bibitem [{\citenamefont {Zhao}\ \emph {et~al.}(2012)\citenamefont {Zhao},
  \citenamefont {Ming}, \citenamefont {Shao}, \citenamefont {Chen},\ and\
  \citenamefont {Wang}}]{Zhao:2012aa}%
  \BibitemOpen
  \bibfield  {author} {\bibinfo {author} {\bibfnamefont {L.}~\bibnamefont
  {Zhao}}, \bibinfo {author} {\bibfnamefont {T.}~\bibnamefont {Ming}}, \bibinfo
  {author} {\bibfnamefont {L.}~\bibnamefont {Shao}}, \bibinfo {author}
  {\bibfnamefont {H.}~\bibnamefont {Chen}}, \ and\ \bibinfo {author}
  {\bibfnamefont {J.}~\bibnamefont {Wang}},\ }\href {\doibase
  10.1021/jp300916a} {\bibfield  {journal} {\bibinfo  {journal} {J. Chem. Phys.
  C}\ }\textbf {\bibinfo {volume} {116}},\ \bibinfo {pages} {8287} (\bibinfo
  {year} {2012})}\BibitemShut {NoStop}%
\bibitem [{\citenamefont {Angioni}\ \emph {et~al.}(2013)\citenamefont
  {Angioni}, \citenamefont {Corni},\ and\ \citenamefont
  {Mennucci}}]{C2CP44010E}%
  \BibitemOpen
  \bibfield  {author} {\bibinfo {author} {\bibfnamefont {A.}~\bibnamefont
  {Angioni}}, \bibinfo {author} {\bibfnamefont {S.}~\bibnamefont {Corni}}, \
  and\ \bibinfo {author} {\bibfnamefont {B.}~\bibnamefont {Mennucci}},\ }\href
  {\doibase 10.1039/C2CP44010E} {\bibfield  {journal} {\bibinfo  {journal}
  {Phys. Chem. Chem. Phys.}\ }\textbf {\bibinfo {volume} {15}},\ \bibinfo
  {pages} {3294} (\bibinfo {year} {2013})}\BibitemShut {NoStop}%
\bibitem [{\citenamefont {Karanikolas}\ \emph {et~al.}(2014)\citenamefont
  {Karanikolas}, \citenamefont {Marocico},\ and\ \citenamefont
  {Bradley}}]{Karanikolas:2014aa}%
  \BibitemOpen
  \bibfield  {author} {\bibinfo {author} {\bibfnamefont {V.}~\bibnamefont
  {Karanikolas}}, \bibinfo {author} {\bibfnamefont {C.~A.}\ \bibnamefont
  {Marocico}}, \ and\ \bibinfo {author} {\bibfnamefont {A.~L.}\ \bibnamefont
  {Bradley}},\ }\href {http://link.aps.org/doi/10.1103/PhysRevA.89.063817}
  {\bibfield  {journal} {\bibinfo  {journal} {Phys. Rev. A}\ }\textbf {\bibinfo
  {volume} {89}},\ \bibinfo {pages} {063817} (\bibinfo {year}
  {2014})}\BibitemShut {NoStop}%
\bibitem [{\citenamefont {Chou}\ and\ \citenamefont
  {Dennis}(2015)}]{chou2015forster}%
  \BibitemOpen
  \bibfield  {author} {\bibinfo {author} {\bibfnamefont {K.~F.}\ \bibnamefont
  {Chou}}\ and\ \bibinfo {author} {\bibfnamefont {A.~M.}\ \bibnamefont
  {Dennis}},\ }\href@noop {} {\bibfield  {journal} {\bibinfo  {journal}
  {Sensors}\ }\textbf {\bibinfo {volume} {15}},\ \bibinfo {pages} {13288}
  (\bibinfo {year} {2015})}\BibitemShut {NoStop}%
\bibitem [{\citenamefont {Marocico}\ \emph {et~al.}(2016)\citenamefont
  {Marocico}, \citenamefont {Zhang},\ and\ \citenamefont {Bradley}}]{4939206}%
  \BibitemOpen
  \bibfield  {author} {\bibinfo {author} {\bibfnamefont {C.~A.}\ \bibnamefont
  {Marocico}}, \bibinfo {author} {\bibfnamefont {X.}~\bibnamefont {Zhang}}, \
  and\ \bibinfo {author} {\bibfnamefont {A.~L.}\ \bibnamefont {Bradley}},\
  }\href@noop {} {\bibfield  {journal} {\bibinfo  {journal} {J. Chem. Phys.}\
  }\textbf {\bibinfo {volume} {144}},\ \bibinfo {eid} {024108} (\bibinfo {year}
  {2016})}\BibitemShut {NoStop}%
\bibitem [{\citenamefont {Kucherenko}\ \emph {et~al.}(2015)\citenamefont
  {Kucherenko}, \citenamefont {Stepanov},\ and\ \citenamefont
  {Kruchinin}}]{Kucherenko2015}%
  \BibitemOpen
  \bibfield  {author} {\bibinfo {author} {\bibfnamefont {M.~G.}\ \bibnamefont
  {Kucherenko}}, \bibinfo {author} {\bibfnamefont {V.~N.}\ \bibnamefont
  {Stepanov}}, \ and\ \bibinfo {author} {\bibfnamefont {N.~Y.}\ \bibnamefont
  {Kruchinin}},\ }\href {\doibase 10.1134/S0030400X15010154} {\bibfield
  {journal} {\bibinfo  {journal} {Opt. Spectrosc+}\ }\textbf {\bibinfo {volume}
  {118}},\ \bibinfo {pages} {103} (\bibinfo {year} {2015})}\BibitemShut
  {NoStop}%
\bibitem [{\citenamefont {Pustovit}\ \emph {et~al.}(2014)\citenamefont
  {Pustovit}, \citenamefont {Urbas},\ and\ \citenamefont
  {Shahbazyan}}]{2040-8986-16-11-114015}%
  \BibitemOpen
  \bibfield  {author} {\bibinfo {author} {\bibfnamefont {V.~N.}\ \bibnamefont
  {Pustovit}}, \bibinfo {author} {\bibfnamefont {A.~M.}\ \bibnamefont {Urbas}},
  \ and\ \bibinfo {author} {\bibfnamefont {T.~V.}\ \bibnamefont {Shahbazyan}},\
  }\href {http://stacks.iop.org/2040-8986/16/i=11/a=114015} {\bibfield
  {journal} {\bibinfo  {journal} {J. Opt.}\ }\textbf {\bibinfo {volume} {16}},\
  \bibinfo {pages} {114015} (\bibinfo {year} {2014})}\BibitemShut {NoStop}%
\bibitem [{\citenamefont {Roslyak}\ \emph {et~al.}(2014)\citenamefont
  {Roslyak}, \citenamefont {Cherqui}, \citenamefont {Dunlap},\ and\
  \citenamefont {Piryatinski}}]{Roslyak:2014aa}%
  \BibitemOpen
  \bibfield  {author} {\bibinfo {author} {\bibfnamefont {O.}~\bibnamefont
  {Roslyak}}, \bibinfo {author} {\bibfnamefont {C.}~\bibnamefont {Cherqui}},
  \bibinfo {author} {\bibfnamefont {D.~H.}\ \bibnamefont {Dunlap}}, \ and\
  \bibinfo {author} {\bibfnamefont {A.}~\bibnamefont {Piryatinski}},\ }\href
  {\doibase 10.1021/jp501144s} {\bibfield  {journal} {\bibinfo  {journal} {J.
  Chem. Phys. B}\ }\textbf {\bibinfo {volume} {118}},\ \bibinfo {pages} {8070}
  (\bibinfo {year} {2014})}\BibitemShut {NoStop}%
\bibitem [{\citenamefont {Barnes}\ \emph {et~al.}(2003)\citenamefont {Barnes},
  \citenamefont {Dereux},\ and\ \citenamefont {Ebbesen}}]{Barnes:2003aa}%
  \BibitemOpen
  \bibfield  {author} {\bibinfo {author} {\bibfnamefont {W.~L.}\ \bibnamefont
  {Barnes}}, \bibinfo {author} {\bibfnamefont {A.}~\bibnamefont {Dereux}}, \
  and\ \bibinfo {author} {\bibfnamefont {T.~W.}\ \bibnamefont {Ebbesen}},\
  }\href {http://dx.doi.org/10.1038/nature01937} {\bibfield  {journal}
  {\bibinfo  {journal} {Nature}\ }\textbf {\bibinfo {volume} {424}},\ \bibinfo
  {pages} {824} (\bibinfo {year} {2003})}\BibitemShut {NoStop}%
\bibitem [{\citenamefont {Maier}\ \emph {et~al.}(2003)\citenamefont {Maier},
  \citenamefont {Kik}, \citenamefont {Atwater}, \citenamefont {Meltzer},
  \citenamefont {Harel}, \citenamefont {Koel},\ and\ \citenamefont
  {Requicha}}]{Maier:2003aa}%
  \BibitemOpen
  \bibfield  {author} {\bibinfo {author} {\bibfnamefont {S.~A.}\ \bibnamefont
  {Maier}}, \bibinfo {author} {\bibfnamefont {P.~G.}\ \bibnamefont {Kik}},
  \bibinfo {author} {\bibfnamefont {H.~A.}\ \bibnamefont {Atwater}}, \bibinfo
  {author} {\bibfnamefont {S.}~\bibnamefont {Meltzer}}, \bibinfo {author}
  {\bibfnamefont {E.}~\bibnamefont {Harel}}, \bibinfo {author} {\bibfnamefont
  {B.~E.}\ \bibnamefont {Koel}}, \ and\ \bibinfo {author} {\bibfnamefont
  {A.~A.~G.}\ \bibnamefont {Requicha}},\ }\href
  {http://dx.doi.org/10.1038/nmat852} {\bibfield  {journal} {\bibinfo
  {journal} {Nat Mater}\ }\textbf {\bibinfo {volume} {2}},\ \bibinfo {pages}
  {229} (\bibinfo {year} {2003})}\BibitemShut {NoStop}%
\bibitem [{\citenamefont {Zia}\ \emph {et~al.}(2004)\citenamefont {Zia},
  \citenamefont {Selker}, \citenamefont {Catrysse},\ and\ \citenamefont
  {Brongersma}}]{Zia:04}%
  \BibitemOpen
  \bibfield  {author} {\bibinfo {author} {\bibfnamefont {R.}~\bibnamefont
  {Zia}}, \bibinfo {author} {\bibfnamefont {M.~D.}\ \bibnamefont {Selker}},
  \bibinfo {author} {\bibfnamefont {P.~B.}\ \bibnamefont {Catrysse}}, \ and\
  \bibinfo {author} {\bibfnamefont {M.~L.}\ \bibnamefont {Brongersma}},\ }\href
  {\doibase 10.1364/JOSAA.21.002442} {\bibfield  {journal} {\bibinfo  {journal}
  {J. Opt. Soc. Am. A}\ }\textbf {\bibinfo {volume} {21}},\ \bibinfo {pages}
  {2442} (\bibinfo {year} {2004})}\BibitemShut {NoStop}%
\bibitem [{\citenamefont {Oulton}\ \emph {et~al.}(2008)\citenamefont {Oulton},
  \citenamefont {Sorger}, \citenamefont {Genov}, \citenamefont {Pile},\ and\
  \citenamefont {Zhang}}]{Oulton:2008aa}%
  \BibitemOpen
  \bibfield  {author} {\bibinfo {author} {\bibfnamefont {R.~F.}\ \bibnamefont
  {Oulton}}, \bibinfo {author} {\bibfnamefont {V.~J.}\ \bibnamefont {Sorger}},
  \bibinfo {author} {\bibfnamefont {D.~A.}\ \bibnamefont {Genov}}, \bibinfo
  {author} {\bibfnamefont {D.~F.~P.}\ \bibnamefont {Pile}}, \ and\ \bibinfo
  {author} {\bibfnamefont {X.}~\bibnamefont {Zhang}},\ }\href
  {http://dx.doi.org/10.1038/nphoton.2008.131} {\bibfield  {journal} {\bibinfo
  {journal} {Nat Photon}\ }\textbf {\bibinfo {volume} {2}},\ \bibinfo {pages}
  {496} (\bibinfo {year} {2008})}\BibitemShut {NoStop}%
\bibitem [{\citenamefont {Grandidier}\ \emph {et~al.}(2009)\citenamefont
  {Grandidier}, \citenamefont {des Francs}, \citenamefont {Massenot},
  \citenamefont {Bouhelier}, \citenamefont {Markey}, \citenamefont {Weeber},
  \citenamefont {Finot},\ and\ \citenamefont {Dereux}}]{Grandidier:2009aa}%
  \BibitemOpen
  \bibfield  {author} {\bibinfo {author} {\bibfnamefont {J.}~\bibnamefont
  {Grandidier}}, \bibinfo {author} {\bibfnamefont {G.~C.}\ \bibnamefont {des
  Francs}}, \bibinfo {author} {\bibfnamefont {S.}~\bibnamefont {Massenot}},
  \bibinfo {author} {\bibfnamefont {A.}~\bibnamefont {Bouhelier}}, \bibinfo
  {author} {\bibfnamefont {L.}~\bibnamefont {Markey}}, \bibinfo {author}
  {\bibfnamefont {J.-C.}\ \bibnamefont {Weeber}}, \bibinfo {author}
  {\bibfnamefont {C.}~\bibnamefont {Finot}}, \ and\ \bibinfo {author}
  {\bibfnamefont {A.}~\bibnamefont {Dereux}},\ }\href {\doibase
  10.1021/nl901314u} {\bibfield  {journal} {\bibinfo  {journal} {Nano Lett.}\
  }\textbf {\bibinfo {volume} {9}},\ \bibinfo {pages} {2935} (\bibinfo {year}
  {2009})}\BibitemShut {NoStop}%
\bibitem [{\citenamefont {Sorger}\ \emph {et~al.}(2011)\citenamefont {Sorger},
  \citenamefont {Ye}, \citenamefont {Oulton}, \citenamefont {Wang},
  \citenamefont {Bartal}, \citenamefont {Yin},\ and\ \citenamefont
  {Zhang}}]{Sorger:2011aa}%
  \BibitemOpen
  \bibfield  {author} {\bibinfo {author} {\bibfnamefont {V.~J.}\ \bibnamefont
  {Sorger}}, \bibinfo {author} {\bibfnamefont {Z.}~\bibnamefont {Ye}}, \bibinfo
  {author} {\bibfnamefont {R.~F.}\ \bibnamefont {Oulton}}, \bibinfo {author}
  {\bibfnamefont {Y.}~\bibnamefont {Wang}}, \bibinfo {author} {\bibfnamefont
  {G.}~\bibnamefont {Bartal}}, \bibinfo {author} {\bibfnamefont
  {X.}~\bibnamefont {Yin}}, \ and\ \bibinfo {author} {\bibfnamefont
  {X.}~\bibnamefont {Zhang}},\ }\href {http://dx.doi.org/10.1038/ncomms1315}
  {\bibfield  {journal} {\bibinfo  {journal} {Nat Commun}\ }\textbf {\bibinfo
  {volume} {2}},\ \bibinfo {pages} {331} (\bibinfo {year} {2011})}\BibitemShut
  {NoStop}%
\bibitem [{\citenamefont {Ellenbogen}\ \emph {et~al.}(2011)\citenamefont
  {Ellenbogen}, \citenamefont {Steinvurzel},\ and\ \citenamefont
  {Crozier}}]{3604014}%
  \BibitemOpen
  \bibfield  {author} {\bibinfo {author} {\bibfnamefont {T.}~\bibnamefont
  {Ellenbogen}}, \bibinfo {author} {\bibfnamefont {P.}~\bibnamefont
  {Steinvurzel}}, \ and\ \bibinfo {author} {\bibfnamefont {K.~B.}\ \bibnamefont
  {Crozier}},\ }\href {\doibase http://dx.doi.org/10.1063/1.3604014} {\bibfield
   {journal} {\bibinfo  {journal} {Appl. Phys. Lett.}\ }\textbf {\bibinfo
  {volume} {98}},\ \bibinfo {eid} {261103} (\bibinfo {year} {2011}),\
  http://dx.doi.org/10.1063/1.3604014}\BibitemShut {NoStop}%
\bibitem [{\citenamefont {Ellenbogen}\ and\ \citenamefont
  {Crozier}(2011)}]{Ellenbogen:2011aa}%
  \BibitemOpen
  \bibfield  {author} {\bibinfo {author} {\bibfnamefont {T.}~\bibnamefont
  {Ellenbogen}}\ and\ \bibinfo {author} {\bibfnamefont {K.~B.}\ \bibnamefont
  {Crozier}},\ }\href {http://link.aps.org/doi/10.1103/PhysRevB.84.161304}
  {\bibfield  {journal} {\bibinfo  {journal} {Phys. Rev. B}\ }\textbf {\bibinfo
  {volume} {84}},\ \bibinfo {pages} {161304} (\bibinfo {year}
  {2011})}\BibitemShut {NoStop}%
\bibitem [{\citenamefont {Paul}\ \emph {et~al.}(2012)\citenamefont {Paul},
  \citenamefont {Solis}, \citenamefont {Bao}, \citenamefont {Chang},
  \citenamefont {Nauert}, \citenamefont {Vidgerman}, \citenamefont {Zubarev},
  \citenamefont {Nordlander},\ and\ \citenamefont {Link}}]{Paul:2012aa}%
  \BibitemOpen
  \bibfield  {author} {\bibinfo {author} {\bibfnamefont {A.}~\bibnamefont
  {Paul}}, \bibinfo {author} {\bibfnamefont {D.}~\bibnamefont {Solis}},
  \bibinfo {author} {\bibfnamefont {K.}~\bibnamefont {Bao}}, \bibinfo {author}
  {\bibfnamefont {W.-S.}\ \bibnamefont {Chang}}, \bibinfo {author}
  {\bibfnamefont {S.}~\bibnamefont {Nauert}}, \bibinfo {author} {\bibfnamefont
  {L.}~\bibnamefont {Vidgerman}}, \bibinfo {author} {\bibfnamefont {E.~R.}\
  \bibnamefont {Zubarev}}, \bibinfo {author} {\bibfnamefont {P.}~\bibnamefont
  {Nordlander}}, \ and\ \bibinfo {author} {\bibfnamefont {S.}~\bibnamefont
  {Link}},\ }\href {\doibase 10.1021/nn3027112} {\bibfield  {journal} {\bibinfo
   {journal} {ACS Nano}\ }\textbf {\bibinfo {volume} {6}},\ \bibinfo {pages}
  {8105} (\bibinfo {year} {2012})}\BibitemShut {NoStop}%
\bibitem [{\citenamefont {Solis}\ \emph {et~al.}(2013)\citenamefont {Solis},
  \citenamefont {Paul}, \citenamefont {Olson}, \citenamefont {Slaughter},
  \citenamefont {Swanglap}, \citenamefont {Chang},\ and\ \citenamefont
  {Link}}]{Solis:2013aa}%
  \BibitemOpen
  \bibfield  {author} {\bibinfo {author} {\bibfnamefont {D.}~\bibnamefont
  {Solis}}, \bibinfo {author} {\bibfnamefont {A.}~\bibnamefont {Paul}},
  \bibinfo {author} {\bibfnamefont {J.}~\bibnamefont {Olson}}, \bibinfo
  {author} {\bibfnamefont {L.~S.}\ \bibnamefont {Slaughter}}, \bibinfo {author}
  {\bibfnamefont {P.}~\bibnamefont {Swanglap}}, \bibinfo {author}
  {\bibfnamefont {W.-S.}\ \bibnamefont {Chang}}, \ and\ \bibinfo {author}
  {\bibfnamefont {S.}~\bibnamefont {Link}},\ }\href {\doibase
  10.1021/nl402358h} {\bibfield  {journal} {\bibinfo  {journal} {Nano Lett.}\
  }\textbf {\bibinfo {volume} {13}},\ \bibinfo {pages} {4779} (\bibinfo {year}
  {2013})}\BibitemShut {NoStop}%
\bibitem [{\citenamefont {Gu}\ \emph {et~al.}(2015)\citenamefont {Gu},
  \citenamefont {Liu}, \citenamefont {Sun}, \citenamefont {Wang}, \citenamefont
  {Lyu}, \citenamefont {Xiao},\ and\ \citenamefont {Song}}]{gu2015photon}%
  \BibitemOpen
  \bibfield  {author} {\bibinfo {author} {\bibfnamefont {Z.}~\bibnamefont
  {Gu}}, \bibinfo {author} {\bibfnamefont {S.}~\bibnamefont {Liu}}, \bibinfo
  {author} {\bibfnamefont {S.}~\bibnamefont {Sun}}, \bibinfo {author}
  {\bibfnamefont {K.}~\bibnamefont {Wang}}, \bibinfo {author} {\bibfnamefont
  {Q.}~\bibnamefont {Lyu}}, \bibinfo {author} {\bibfnamefont {S.}~\bibnamefont
  {Xiao}}, \ and\ \bibinfo {author} {\bibfnamefont {Q.}~\bibnamefont {Song}},\
  }\href@noop {} {\bibfield  {journal} {\bibinfo  {journal} {Scientific
  Reports}\ }\textbf {\bibinfo {volume} {5}} (\bibinfo {year}
  {2015})}\BibitemShut {NoStop}%
\bibitem [{\citenamefont {Carmeli}\ \emph {et~al.}(2015)\citenamefont
  {Carmeli}, \citenamefont {Cohen}, \citenamefont {Heifler}, \citenamefont
  {Lilach}, \citenamefont {Zalevsky}, \citenamefont {Mujica},\ and\
  \citenamefont {Richter}}]{Carmeli:2015aa}%
  \BibitemOpen
  \bibfield  {author} {\bibinfo {author} {\bibfnamefont {I.}~\bibnamefont
  {Carmeli}}, \bibinfo {author} {\bibfnamefont {M.}~\bibnamefont {Cohen}},
  \bibinfo {author} {\bibfnamefont {O.}~\bibnamefont {Heifler}}, \bibinfo
  {author} {\bibfnamefont {Y.}~\bibnamefont {Lilach}}, \bibinfo {author}
  {\bibfnamefont {Z.}~\bibnamefont {Zalevsky}}, \bibinfo {author}
  {\bibfnamefont {V.}~\bibnamefont {Mujica}}, \ and\ \bibinfo {author}
  {\bibfnamefont {S.}~\bibnamefont {Richter}},\ }\href
  {http://dx.doi.org/10.1038/ncomms8334} {\bibfield  {journal} {\bibinfo
  {journal} {Nat Commun}\ }\textbf {\bibinfo {volume} {6}} (\bibinfo {year}
  {2015})}\BibitemShut {NoStop}%
\bibitem [{\citenamefont {Lopata}\ and\ \citenamefont
  {Neuhauser}(2009)}]{3167407}%
  \BibitemOpen
  \bibfield  {author} {\bibinfo {author} {\bibfnamefont {K.}~\bibnamefont
  {Lopata}}\ and\ \bibinfo {author} {\bibfnamefont {D.}~\bibnamefont
  {Neuhauser}},\ }\href
  {http://scitation.aip.org/content/aip/journal/jcp/131/1/10.1063/1.3167407}
  {\bibfield  {journal} {\bibinfo  {journal} {J. Chem. Phys.}\ }\textbf
  {\bibinfo {volume} {131}},\ \bibinfo {eid} {014701} (\bibinfo {year}
  {2009})}\BibitemShut {NoStop}%
\bibitem [{\citenamefont {Arntsen}\ \emph {et~al.}(2011)\citenamefont
  {Arntsen}, \citenamefont {Lopata}, \citenamefont {Wall}, \citenamefont
  {Bartell},\ and\ \citenamefont {Neuhauser}}]{3541820}%
  \BibitemOpen
  \bibfield  {author} {\bibinfo {author} {\bibfnamefont {C.}~\bibnamefont
  {Arntsen}}, \bibinfo {author} {\bibfnamefont {K.}~\bibnamefont {Lopata}},
  \bibinfo {author} {\bibfnamefont {M.~R.}\ \bibnamefont {Wall}}, \bibinfo
  {author} {\bibfnamefont {L.}~\bibnamefont {Bartell}}, \ and\ \bibinfo
  {author} {\bibfnamefont {D.}~\bibnamefont {Neuhauser}},\ }\href
  {http://scitation.aip.org/content/aip/journal/jcp/134/8/10.1063/1.3541820}
  {\bibfield  {journal} {\bibinfo  {journal} {J. Chem. Phys.}\ }\textbf
  {\bibinfo {volume} {134}},\ \bibinfo {eid} {084101} (\bibinfo {year}
  {2011})}\BibitemShut {NoStop}%
\bibitem [{\citenamefont {Sukharev}\ and\ \citenamefont
  {Nitzan}(2011)}]{PhysRevA.84.043802}%
  \BibitemOpen
  \bibfield  {author} {\bibinfo {author} {\bibfnamefont {M.}~\bibnamefont
  {Sukharev}}\ and\ \bibinfo {author} {\bibfnamefont {A.}~\bibnamefont
  {Nitzan}},\ }\href {\doibase 10.1103/PhysRevA.84.043802} {\bibfield
  {journal} {\bibinfo  {journal} {Phys. Rev. A}\ }\textbf {\bibinfo {volume}
  {84}},\ \bibinfo {pages} {043802} (\bibinfo {year} {2011})}\BibitemShut
  {NoStop}%
\bibitem [{\citenamefont {Bonifacio}\ and\ \citenamefont
  {Lugiato}(1975)}]{Bonifacio:1975aa}%
  \BibitemOpen
  \bibfield  {author} {\bibinfo {author} {\bibfnamefont {R.}~\bibnamefont
  {Bonifacio}}\ and\ \bibinfo {author} {\bibfnamefont {L.~A.}\ \bibnamefont
  {Lugiato}},\ }\href {http://link.aps.org/doi/10.1103/PhysRevA.11.1507}
  {\bibfield  {journal} {\bibinfo  {journal} {Phys. Rev. A}\ }\textbf {\bibinfo
  {volume} {11}},\ \bibinfo {pages} {1507} (\bibinfo {year}
  {1975})}\BibitemShut {NoStop}%
\bibitem [{\citenamefont {Bowden}\ and\ \citenamefont
  {Dowling}(1993)}]{PhysRevA.47.1247}%
  \BibitemOpen
  \bibfield  {author} {\bibinfo {author} {\bibfnamefont {C.~M.}\ \bibnamefont
  {Bowden}}\ and\ \bibinfo {author} {\bibfnamefont {J.~P.}\ \bibnamefont
  {Dowling}},\ }\href {\doibase 10.1103/PhysRevA.47.1247} {\bibfield  {journal}
  {\bibinfo  {journal} {Phys. Rev. A}\ }\textbf {\bibinfo {volume} {47}},\
  \bibinfo {pages} {1247} (\bibinfo {year} {1993})}\BibitemShut {NoStop}%
\bibitem [{\citenamefont {Judkins}\ and\ \citenamefont
  {Ziolkowski}(1995)}]{Judkins:95}%
  \BibitemOpen
  \bibfield  {author} {\bibinfo {author} {\bibfnamefont {J.~B.}\ \bibnamefont
  {Judkins}}\ and\ \bibinfo {author} {\bibfnamefont {R.~W.}\ \bibnamefont
  {Ziolkowski}},\ }\href {\doibase 10.1364/JOSAA.12.001974} {\bibfield
  {journal} {\bibinfo  {journal} {J. Opt. Soc. Am. A}\ }\textbf {\bibinfo
  {volume} {12}},\ \bibinfo {pages} {1974} (\bibinfo {year}
  {1995})}\BibitemShut {NoStop}%
\bibitem [{\citenamefont {Hofmann}\ and\ \citenamefont
  {Hess}(1999)}]{Hofmann:1999aa}%
  \BibitemOpen
  \bibfield  {author} {\bibinfo {author} {\bibfnamefont {H.~F.}\ \bibnamefont
  {Hofmann}}\ and\ \bibinfo {author} {\bibfnamefont {O.}~\bibnamefont {Hess}},\
  }\href {http://link.aps.org/doi/10.1103/PhysRevA.59.2342} {\bibfield
  {journal} {\bibinfo  {journal} {Phys. Rev. A}\ }\textbf {\bibinfo {volume}
  {59}},\ \bibinfo {pages} {2342} (\bibinfo {year} {1999})}\BibitemShut
  {NoStop}%
\bibitem [{\citenamefont {Slavcheva}\ \emph {et~al.}(2002)\citenamefont
  {Slavcheva}, \citenamefont {Arnold}, \citenamefont {Wallace},\ and\
  \citenamefont {Ziolkowski}}]{Slavcheva:2002aa}%
  \BibitemOpen
  \bibfield  {author} {\bibinfo {author} {\bibfnamefont {G.}~\bibnamefont
  {Slavcheva}}, \bibinfo {author} {\bibfnamefont {J.~M.}\ \bibnamefont
  {Arnold}}, \bibinfo {author} {\bibfnamefont {I.}~\bibnamefont {Wallace}}, \
  and\ \bibinfo {author} {\bibfnamefont {R.~W.}\ \bibnamefont {Ziolkowski}},\
  }\href {http://link.aps.org/doi/10.1103/PhysRevA.66.063418} {\bibfield
  {journal} {\bibinfo  {journal} {Phys. Rev. A}\ }\textbf {\bibinfo {volume}
  {66}},\ \bibinfo {pages} {063418} (\bibinfo {year} {2002})}\BibitemShut
  {NoStop}%
\bibitem [{\citenamefont {Slavcheva}\ \emph {et~al.}(2004)\citenamefont
  {Slavcheva}, \citenamefont {Arnold},\ and\ \citenamefont
  {Ziolkowski}}]{Slavcheva:2004aa}%
  \BibitemOpen
  \bibfield  {author} {\bibinfo {author} {\bibfnamefont {G.~M.}\ \bibnamefont
  {Slavcheva}}, \bibinfo {author} {\bibfnamefont {J.~M.}\ \bibnamefont
  {Arnold}}, \ and\ \bibinfo {author} {\bibfnamefont {R.~W.}\ \bibnamefont
  {Ziolkowski}},\ }\href {\doibase 10.1109/JSTQE.2004.836023} {\bibfield
  {journal} {\bibinfo  {journal} {IEEE J. Sel. Top. Quant.}\ }\textbf {\bibinfo
  {volume} {10}},\ \bibinfo {pages} {1052} (\bibinfo {year}
  {2004})}\BibitemShut {NoStop}%
\bibitem [{\citenamefont {Fratalocchi}\ \emph {et~al.}(2008)\citenamefont
  {Fratalocchi}, \citenamefont {Conti},\ and\ \citenamefont
  {Ruocco}}]{Fratalocchi:2008aa}%
  \BibitemOpen
  \bibfield  {author} {\bibinfo {author} {\bibfnamefont {A.}~\bibnamefont
  {Fratalocchi}}, \bibinfo {author} {\bibfnamefont {C.}~\bibnamefont {Conti}},
  \ and\ \bibinfo {author} {\bibfnamefont {G.}~\bibnamefont {Ruocco}},\ }\href
  {http://link.aps.org/doi/10.1103/PhysRevA.78.013806} {\bibfield  {journal}
  {\bibinfo  {journal} {Phys. Rev. A}\ }\textbf {\bibinfo {volume} {78}},\
  \bibinfo {pages} {013806} (\bibinfo {year} {2008})}\BibitemShut {NoStop}%
\bibitem [{\citenamefont {Andreasen}\ and\ \citenamefont
  {Cao}(2009)}]{Andreasen:09}%
  \BibitemOpen
  \bibfield  {author} {\bibinfo {author} {\bibfnamefont {J.}~\bibnamefont
  {Andreasen}}\ and\ \bibinfo {author} {\bibfnamefont {H.}~\bibnamefont
  {Cao}},\ }\href {http://jlt.osa.org/abstract.cfm?URI=jlt-27-20-4530}
  {\bibfield  {journal} {\bibinfo  {journal} {J. Lightwave Technol.}\ }\textbf
  {\bibinfo {volume} {27}},\ \bibinfo {pages} {4530} (\bibinfo {year}
  {2009})}\BibitemShut {NoStop}%
\bibitem [{\citenamefont {Sukharev}\ \emph {et~al.}(2014)\citenamefont
  {Sukharev}, \citenamefont {Seideman}, \citenamefont {Gordon}, \citenamefont
  {Salomon},\ and\ \citenamefont {Prior}}]{doi:10.1021/nn4054528}%
  \BibitemOpen
  \bibfield  {author} {\bibinfo {author} {\bibfnamefont {M.}~\bibnamefont
  {Sukharev}}, \bibinfo {author} {\bibfnamefont {T.}~\bibnamefont {Seideman}},
  \bibinfo {author} {\bibfnamefont {R.~J.}\ \bibnamefont {Gordon}}, \bibinfo
  {author} {\bibfnamefont {A.}~\bibnamefont {Salomon}}, \ and\ \bibinfo
  {author} {\bibfnamefont {Y.}~\bibnamefont {Prior}},\ }\href {\doibase
  10.1021/nn4054528} {\bibfield  {journal} {\bibinfo  {journal} {ACS Nano}\
  }\textbf {\bibinfo {volume} {8}},\ \bibinfo {pages} {807} (\bibinfo {year}
  {2014})}\BibitemShut {NoStop}%
\bibitem [{\citenamefont {Sukharev}(2014)}]{jcp_chirps14}%
  \BibitemOpen
  \bibfield  {author} {\bibinfo {author} {\bibfnamefont {M.}~\bibnamefont
  {Sukharev}},\ }\href {\doibase http://dx.doi.org/10.1063/1.4893967}
  {\bibfield  {journal} {\bibinfo  {journal} {J. Chem. Phys.}\ }\textbf
  {\bibinfo {volume} {141}},\ \bibinfo {eid} {084712} (\bibinfo {year}
  {2014})}\BibitemShut {NoStop}%
\bibitem [{\citenamefont {Sukharev}\ \emph {et~al.}(2015)\citenamefont
  {Sukharev}, \citenamefont {Day},\ and\ \citenamefont
  {Pachter}}]{Sukharev:2015aa}%
  \BibitemOpen
  \bibfield  {author} {\bibinfo {author} {\bibfnamefont {M.}~\bibnamefont
  {Sukharev}}, \bibinfo {author} {\bibfnamefont {P.~N.}\ \bibnamefont {Day}}, \
  and\ \bibinfo {author} {\bibfnamefont {R.}~\bibnamefont {Pachter}},\ }\href
  {\doibase 10.1021/acsphotonics.5b00146} {\bibfield  {journal} {\bibinfo
  {journal} {ACS Photon.}\ }\textbf {\bibinfo {volume} {2}},\ \bibinfo {pages}
  {935} (\bibinfo {year} {2015})}\BibitemShut {NoStop}%
\bibitem [{\citenamefont {Blake}\ and\ \citenamefont
  {Sukharev}(2015)}]{Blake:2015aa}%
  \BibitemOpen
  \bibfield  {author} {\bibinfo {author} {\bibfnamefont {A.}~\bibnamefont
  {Blake}}\ and\ \bibinfo {author} {\bibfnamefont {M.}~\bibnamefont
  {Sukharev}},\ }\href {http://link.aps.org/doi/10.1103/PhysRevB.92.035433}
  {\bibfield  {journal} {\bibinfo  {journal} {Phys. Rev. B}\ }\textbf {\bibinfo
  {volume} {92}},\ \bibinfo {pages} {035433} (\bibinfo {year}
  {2015})}\BibitemShut {NoStop}%
\bibitem [{\citenamefont {T{\"o}rm{\"a}}\ and\ \citenamefont
  {Barnes}(2015)}]{0034-4885-78-1-013901}%
  \BibitemOpen
  \bibfield  {author} {\bibinfo {author} {\bibfnamefont {P.}~\bibnamefont
  {T{\"o}rm{\"a}}}\ and\ \bibinfo {author} {\bibfnamefont {W.~L.}\ \bibnamefont
  {Barnes}},\ }\href {http://stacks.iop.org/0034-4885/78/i=1/a=013901}
  {\bibfield  {journal} {\bibinfo  {journal} {Rep. Prog. Phys.}\ }\textbf
  {\bibinfo {volume} {78}},\ \bibinfo {pages} {013901} (\bibinfo {year}
  {2015})}\BibitemShut {NoStop}%
\bibitem [{\citenamefont {Benner}\ \emph {et~al.}(2013)\citenamefont {Benner},
  \citenamefont {Boneberg}, \citenamefont {N{\"u}rnberger}, \citenamefont
  {Ghafoori}, \citenamefont {Leiderer},\ and\ \citenamefont
  {Scheer}}]{1367-2630-15-11-113014}%
  \BibitemOpen
  \bibfield  {author} {\bibinfo {author} {\bibfnamefont {D.}~\bibnamefont
  {Benner}}, \bibinfo {author} {\bibfnamefont {J.}~\bibnamefont {Boneberg}},
  \bibinfo {author} {\bibfnamefont {P.}~\bibnamefont {N{\"u}rnberger}},
  \bibinfo {author} {\bibfnamefont {G.}~\bibnamefont {Ghafoori}}, \bibinfo
  {author} {\bibfnamefont {P.}~\bibnamefont {Leiderer}}, \ and\ \bibinfo
  {author} {\bibfnamefont {E.}~\bibnamefont {Scheer}},\ }\href
  {http://stacks.iop.org/1367-2630/15/i=11/a=113014} {\bibfield  {journal}
  {\bibinfo  {journal} {New J. Phys.}\ }\textbf {\bibinfo {volume} {15}},\
  \bibinfo {pages} {113014} (\bibinfo {year} {2013})}\BibitemShut {NoStop}%
\bibitem [{\citenamefont {Benner}\ \emph {et~al.}(2014)\citenamefont {Benner},
  \citenamefont {Boneberg}, \citenamefont {N{\"u}rnberger}, \citenamefont
  {Waitz}, \citenamefont {Leiderer},\ and\ \citenamefont
  {Scheer}}]{Benner:2014aa}%
  \BibitemOpen
  \bibfield  {author} {\bibinfo {author} {\bibfnamefont {D.}~\bibnamefont
  {Benner}}, \bibinfo {author} {\bibfnamefont {J.}~\bibnamefont {Boneberg}},
  \bibinfo {author} {\bibfnamefont {P.}~\bibnamefont {N{\"u}rnberger}},
  \bibinfo {author} {\bibfnamefont {R.}~\bibnamefont {Waitz}}, \bibinfo
  {author} {\bibfnamefont {P.}~\bibnamefont {Leiderer}}, \ and\ \bibinfo
  {author} {\bibfnamefont {E.}~\bibnamefont {Scheer}},\ }\href {\doibase
  10.1021/nl502165y} {\bibfield  {journal} {\bibinfo  {journal} {Nano Lett.}\
  }\textbf {\bibinfo {volume} {14}},\ \bibinfo {pages} {5218} (\bibinfo {year}
  {2014})}\BibitemShut {NoStop}%
\bibitem [{\citenamefont {Sadeghi}(2012)}]{4767653}%
  \BibitemOpen
  \bibfield  {author} {\bibinfo {author} {\bibfnamefont {S.~M.}\ \bibnamefont
  {Sadeghi}},\ }\href {\doibase 10.1063/1.4767653} {\bibfield  {journal}
  {\bibinfo  {journal} {Appl. Phys. Lett.}\ }\textbf {\bibinfo {volume}
  {101}},\ \bibinfo {eid} {213102} (\bibinfo {year} {2012}),\
  10.1063/1.4767653}\BibitemShut {NoStop}%
\bibitem [{\citenamefont {Norton}\ and\ \citenamefont
  {Vo~Dinh}(2008)}]{doi:10.1117/1.3001731}%
  \BibitemOpen
  \bibfield  {author} {\bibinfo {author} {\bibfnamefont {S.}~\bibnamefont
  {Norton}}\ and\ \bibinfo {author} {\bibfnamefont {T.}~\bibnamefont
  {Vo~Dinh}},\ }\href {\doibase 10.1117/1.3001731} {\bibfield  {journal}
  {\bibinfo  {journal} {J. Nanophotonics}\ }\textbf {\bibinfo {volume} {2}},\
  \bibinfo {pages} {029501} (\bibinfo {year} {2008})}\BibitemShut {NoStop}%
\bibitem [{\citenamefont {Gray}\ and\ \citenamefont
  {Kupka}(2003)}]{PhysRevB.68.045415}%
  \BibitemOpen
  \bibfield  {author} {\bibinfo {author} {\bibfnamefont {S.~K.}\ \bibnamefont
  {Gray}}\ and\ \bibinfo {author} {\bibfnamefont {T.}~\bibnamefont {Kupka}},\
  }\href {\doibase 10.1103/PhysRevB.68.045415} {\bibfield  {journal} {\bibinfo
  {journal} {Phys. Rev. B}\ }\textbf {\bibinfo {volume} {68}},\ \bibinfo
  {pages} {045415} (\bibinfo {year} {2003})}\BibitemShut {NoStop}%
\bibitem [{\citenamefont {Siegman}(1986)}]{siegman1986university}%
  \BibitemOpen
  \bibfield  {author} {\bibinfo {author} {\bibfnamefont {A.~E.}\ \bibnamefont
  {Siegman}},\ }\href@noop {} {\emph {\bibinfo {title} {Lasers}}}\ (\bibinfo
  {publisher} {University Science Books, Mill Valley, CA},\ \bibinfo {year}
  {1986})\BibitemShut {NoStop}%
\bibitem [{\citenamefont {Puthumpally-Joseph}\ \emph
  {et~al.}(2015)\citenamefont {Puthumpally-Joseph}, \citenamefont {Atabek},
  \citenamefont {Sukharev},\ and\ \citenamefont
  {Charron}}]{PhysRevA.91.043835}%
  \BibitemOpen
  \bibfield  {author} {\bibinfo {author} {\bibfnamefont {R.}~\bibnamefont
  {Puthumpally-Joseph}}, \bibinfo {author} {\bibfnamefont {O.}~\bibnamefont
  {Atabek}}, \bibinfo {author} {\bibfnamefont {M.}~\bibnamefont {Sukharev}}, \
  and\ \bibinfo {author} {\bibfnamefont {E.}~\bibnamefont {Charron}},\ }\href
  {\doibase 10.1103/PhysRevA.91.043835} {\bibfield  {journal} {\bibinfo
  {journal} {Phys. Rev. A}\ }\textbf {\bibinfo {volume} {91}},\ \bibinfo
  {pages} {043835} (\bibinfo {year} {2015})}\BibitemShut {NoStop}%
\bibitem [{\citenamefont {Taflove}\ and\ \citenamefont
  {Hagness}(2005)}]{taflove2005computational}%
  \BibitemOpen
  \bibfield  {author} {\bibinfo {author} {\bibfnamefont {A.}~\bibnamefont
  {Taflove}}\ and\ \bibinfo {author} {\bibfnamefont {S.}~\bibnamefont
  {Hagness}},\ }\href@noop {} {\emph {\bibinfo {title} {Computational
  Electrodynamics: The Finite-Difference Time-Domain Method}}},\ \bibinfo
  {edition} {3rd}\ ed.\ (\bibinfo  {publisher} {Artech House, Boston},\
  \bibinfo {year} {2005})\BibitemShut {NoStop}%
\bibitem [{\citenamefont {Roden}\ and\ \citenamefont
  {Gedney}(2000)}]{CPML_paper}%
  \BibitemOpen
  \bibfield  {author} {\bibinfo {author} {\bibfnamefont {J.~A.}\ \bibnamefont
  {Roden}}\ and\ \bibinfo {author} {\bibfnamefont {S.~D.}\ \bibnamefont
  {Gedney}},\ }\href {\doibase
  10.1002/1098-2760(20001205)27:5<334::AID-MOP14>3.0.CO;2-A} {\bibfield
  {journal} {\bibinfo  {journal} {Microw. Opt. Techn. Lett.}\ }\textbf
  {\bibinfo {volume} {27}},\ \bibinfo {pages} {334} (\bibinfo {year}
  {2000})}\BibitemShut {NoStop}%
\bibitem [{\citenamefont {Liz-Marz{\'a}n}(2006)}]{doi:10.1021/la0513353}%
  \BibitemOpen
  \bibfield  {author} {\bibinfo {author} {\bibfnamefont {L.~M.}\ \bibnamefont
  {Liz-Marz{\'a}n}},\ }\href {\doibase 10.1021/la0513353} {\bibfield  {journal}
  {\bibinfo  {journal} {Langmuir}\ }\textbf {\bibinfo {volume} {22}},\ \bibinfo
  {pages} {32} (\bibinfo {year} {2006})}\BibitemShut {NoStop}%
\end{thebibliography}
\end{document}